# Identifying dominant recombination mechanisms in perovskite solar cells by measuring the transient ideality factor

Phil Calado[1†*], Dan Burkitt[2†], Jizhong Yao[1†], Joel Troughton[2], Trystan M. Watson[2], Matt J. Carnie[2], Andrew M. Telford[1], Brian C. O'Regan[3], Jenny Nelson[1], Piers R. F. Barnes*[1]

1. Department of Physics, Blackett Laboratory, Imperial College London SW7 2AZ London, UK
2. SPECIFIC, College of Engineering, Swansea University, Baglan Bay Innovation and Knowledge Centre, Central Avenue, Baglan, SA12 7AX, UK.
3. Sunlight Scientific, Berkeley, CA, USA

†These authors contributed equally to this work

*p.calado13@imperial.ac.uk
*piers.barnes@imperial.ac.uk

**Abstract**
The ideality factor determined by measuring the open circuit voltage ($V_{OC}$) as function of light intensity is often used as a means to identify the dominant recombination mechanism in solar cells. However, applying this 'Suns-$V_{OC}$' technique to perovskite cells is problematic since the $V_{OC}$ evolves with time in a way which depends on the previously applied bias ($V_{pre}$), the light intensity, and the device architecture/processing. Here we show that the dominant recombination mechanism in two structurally similar $CH_3NH_3PbI_3$ devices containing either mesoporous $Al_2O_3$ or $TiO_2$ layers can be identified from the signature of the transient ideality factor following application of a forward bias, $V_{pre}$, to the device in the dark. The transient ideality factor, is measured by monitoring the evolution of $V_{OC}$ as a function of time at different light intensities. The initial values of ideality found using this technique were consistent with corresponding photoluminescence *vs* intensity as well as electroluminescence *vs* current density measurements. Time-dependent simulations of the measurement on modelled devices, which include the effects of mobile ionic charge, show that Shockley Read Hall (SRH) recombination through deep traps at the charge collection interfaces is dominant in both devices. Using superimposed transient photovoltage measurements of the evolving $V_{OC}$ on bifacial devices we further show that the charge collection interface extends throughout the mesoporous $TiO_2$ layer, consistent with a transient ideality signature corresponding to SRH recombination in the bulk of the film. This information could not be inferred from an ideality factor determined from the steady-state $V_{OC}$ values alone. We anticipate that the method we have developed will be useful for identifying performance bottlenecks in new variants of perovskite solar cells by comparison with the transient ideality signatures we have predicted for a range of possible recombination schemes.

**Introduction**

Identifying the dominant recombination mechanisms in hybrid perovskite solar cells is necessary to rationally optimise device architectures for performance and stability. The 'light' ideality factor, $n_{id}$ (also known as the 'quality factor'), determined from the open circuit voltage ($V_{OC}$) dependence on light intensity, has classically been employed as a means to identify the dominant recombination mechanism in silicon and organic absorber solar cells.[1–3] However, applying this technique to perovskite cells is problematic since these devices often show hysteresis in their current-voltage (J-V) characteristics, with an associated slow transient evolution of the $V_{OC}$.[4] This evolution depends on the light and voltage bias device preconditioning. Accordingly, for many perovskite cells, ideality factors derived from these measurements depend on the time at which the $V_{OC}$ is evaluated after illuminating the device as well as the state the cell was in prior to illumination.

Pockett *et al.* circumvented the issue of $V_{OC}$ evolution by using open circuit voltage values after the cell has reached steady-state as function of light intensity to determine ideality factors; a number of devices in their study showed anomalously high values of $n_{id} > 5$.[5] Tress *et al.* conducted light intensity scanning sweeps in both increasing and decreasing directions and took the $V_{OC}$ value after 3 seconds as a compromise between stabilisation of the $V_{OC}$ and irreversible degradation of devices.[6] Optical measurements of the ideality factor on perovskite monolayers and heterojunctions have also been made under steady-state conditions.[7] To date, however, a detailed understanding of the time-dependent ideality factor in perovskite solar cells has remained elusive.

There is now considerable evidence to indicate that J-V hysteresis in hybrid perovskite devices results from a combination of the effects of redistribution of mobile ionic defects and high rates of recombination at the perovskite-transport layer interfaces.[4,8–12] Consequently, it is tempting to speculate that the dominant recombination mechanism in a device evolves with the $V_{OC}$ evolution, or that the recombination mechanism at steady-state could be a function of light intensity. The results presented herein show that such changes are not required to explain much of the behaviour observed when examining the ideality factor of perovskite solar cells.

In this study we present a method for determining the transient ideality factor, $n_{id}(t)$, of a device as it evolves from an initial state, defined by a voltage preconditioning protocol in the dark, towards a steady-state value after illumination. The measurement, made by monitoring the evolution of $V_{OC}$ for different light intensities, (which we will refer to as a 'Transient Suns-$V_{OC}$' measurement) is presented for two different device architectures which contain either a mesoporous TiO$_2$ or Al$_2$O$_3$ layer. While a mesoporous TiO$_2$ layer is expected to contribute to electron collection and possibly recombination, the electrically insulating nature of mesoporous Al$_2$O$_3$ should not allow charge transfer across its interface. Comparing the two architectures thus provides a means to control the details of possible recombination mechanisms. To cross-check the Transient Suns-$V_{OC}$ measurement we also present corresponding instantaneous ideality factor measurements determined from electroluminescence *vs* current density (J-EL) and photoluminescence *vs* light intensity (Suns-PL). These observations are supplemented with fast, wavelength dependent, transient photovoltage measurements superimposed on the evolving $V_{OC}$ enabling us to probe the behaviour of charge carriers near the interfaces. In order to interpret the experimental findings we perform drift-diffusion device simulations describing the time-dependent evolution of the free electron, hole and ionic defect concentration profiles using an existing open source simulation tool, Driftfusion.[4,13] The results show that devices with different recombination schemes exhibit unique $n_{id}(t)$ 'signatures', determined by the change in electron and hole population overlap resulting from the screening of internal electric fields by mobile ionic charge. Following forward bias the initial and final values of the $n_{id}(t)$ can be related to the established quasi-zero dimensional model, enabling the dominant recombination mechanism to be inferred in devices. It is not possible to infer this information from the steady-state ideality factor alone.

**Background**

*The origins of ideality factors*

We first examine the origins of the ideality factor without explicitly considering the influence of mobile ions. The *J-V* characteristics of a solar cell can be described by the non-ideal photodiode equation in which the exponent contains an ideality factor to account for the deviation of the current from the value predicted by Boltzmann statistics:

$$J = J_{SC} - J_0 \left( \exp\left(\frac{qV}{n_{id}k_B T}\right) - 1 \right) \qquad (1)$$

where *J* is the current density, $J_{SC}$ is the short-circuit current density, $J_0$ is the dark saturation current density, *V* is the cell voltage, $k_B$ is Boltzmann's constant and *T* is the temperature. Conventionally, the ideality factor is measured either from the slope of the dark ln(*J*) *vs V* characteristics (where $J_{SC}$ = 0) or, in order to avoid the effects of series resistance, by measuring the open circuit voltage as a function of light intensity, or incident photon flux *φ* (known as a 'Suns-$V_{OC}$' measurement). Using equation 1 and assuming *φ* ∝ $J_{SC}$, the ideality factor can be obtained from the $V_{OC}$ *vs* ln(*φ*) curve:

$$\ln \phi = \frac{qV_{OC}}{n_{id}k_B T} + constant \qquad (2)$$

At open circuit the rate of charge carrier generation is equal to the rate of recombination, *U*, and thus proportional to *φ*, allowing the ideality factor to be expressed in terms of the recombination rate and $V_{OC}$:

$$n_{id} = \frac{q}{k_B T} \frac{dV_{OC}}{d \ln U} \qquad (3)$$

Ideality factors are typically related to recombination mechanisms and their associated reaction orders using a zero-dimensional theoretical framework. First, a relationship between the quasi Fermi-level splitting, $\Delta E_F$ (which is equivalent to $qV_{OC}$ in zero dimensions) and free electron (or hole) carrier density *n* (or *p*) is defined such that:

$$\Delta E_F = k_B T \ln\left(\frac{np}{n_i^2}\right) = k_B T \ln\left(\frac{n^\beta}{n_i^2}\right) \qquad (4)$$

where $n_i$ is the equilibrium charge carrier density and *β* is a parameter defining the relationship between the charge carrier density and the perturbation of the quasi Fermi-levels from equilibrium $E_{F0}$. If *n* ≈ *p*, then *np* → $n^2$ and *β* → 2. In this case $\Delta E_F$ is split equally between the individual electron ($E_{Fn}$) and hole ($E_{Fp}$) quasi Fermi-levels (see Figure 1a).

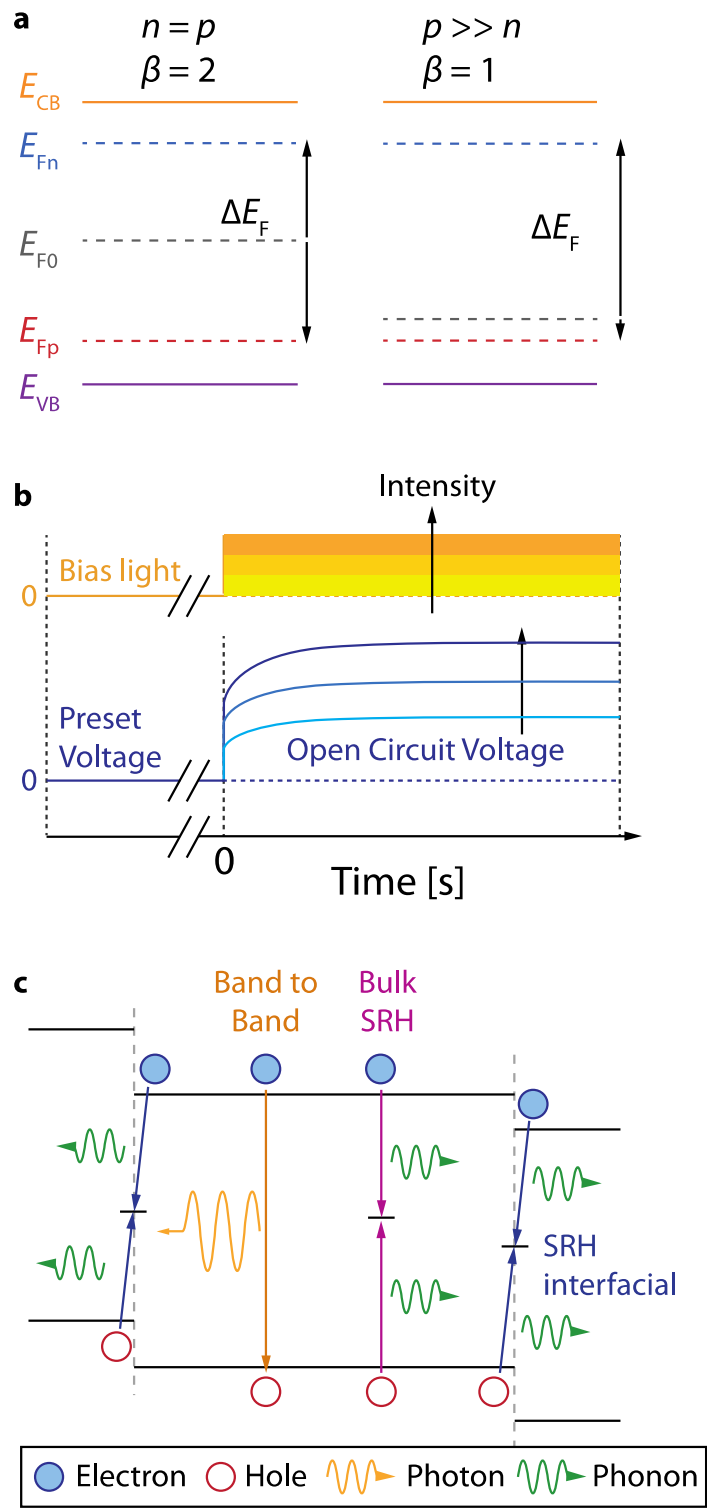

**Figure 1. Schematics of quasi Fermi level splitting, transient Suns-$V_{OC}$ experimental timeline and recombination mechanisms.** (**a**) Schematic of idealised quasi Fermi level splitting $\Delta E_F$ in a zero-dimensional model. Where $n = p$, the Fermi levels split with equal magnitude from equilibrium $E_{F0}$ under bias. Where one carrier is in excess ($p \gg n$ is shown) the shift in the minority carrier Fermi energy dominates the change in $\Delta E_F$. (**b**) Transient Suns-$V_{OC}$ experimental timeline: the cell is preconditioned in the dark at a preset voltage $V_{pre}$ for > 60 s. At $t = 0$ the bias light is switched on and the cell is simultaneously switched to open circuit. The $V_{OC}$ is subsequently measured for > 30 s. The protocol is cycled at increasing bias light intensity and the ideality factor as a function of time is calculated. (**c**) Schematic illustrating different possible recombination pathways within a device.

Where a majority carrier type exists, $β → 1$. This can be shown by expressing the carrier concentrations in terms of their change from equilibrium values ($n_0$ and $p_0$) such that $n = Δn + n_0$ and $p = Δp + p_0$. If $p_0 >> Δp = Δn >> n_0$ then the charge density ratio in the logarithm of equation 4 can be written:

$$\frac{np}{n_i^2} = \frac{(n_0+Δn)(p_0+Δp)}{n_i^2} ≈ \frac{np_0}{n_i^2} \tag{5}$$

In this case it is apparent that $ΔE_F$ will be almost exclusively influenced by the changes in the minority carrier, $n$. Under the assumption that a single recombination reaction order, $γ$, dominates the behaviour of the device such that $U = kn^γ$ (where $k$ is the order dependent recombination rate coefficient) and by substituting equation 4 into equation 3, the relationship between $n_{id}$, $β$, and $γ$ can be derived (see supporting information, Note 1) such that:

$$n_{id} = \frac{β}{γ} \tag{6}$$

The ideality factor is therefore given by a ratio of the relationship between carrier density and quasi Fermi-level splitting, $β$, and the dominant recombination reaction order, $γ$. Table 1 summarises these quantities for different recombination mechansims, trap energies and relative carrier population overlap cases. The derivation of the reaction orders corresponding to band-to-band recombination or trap mediated Schockley-Read-Hall (SRH)[14] recombination *via* different trap depths are given in the supporting information, Note 2. It is apparent from Table 1 that in the limiting cases, $γ$ and $β$ often cancel, resulting in an ideality of $n_{id} = 1$. SRH recombiantion via mid-gap trap energies is the only mechanism listed that results in a value of $n_{id} = 2$.

**Table 1.** Carrier density to $ΔE_F$ relationship $β$, recombination reaction order $γ$, and ideality factor, $n_{id} = β/γ$, for different recombination mechanisms, trap energies and carrier population overlaps. The derivation of the different values of $γ$ are given in the supporting information Note 2. Note that in the case of shallow traps, the values of $β$, $γ$ and $n_{id}$ tend towards those for band-to-band recombination. *$γ = 1$ (and thus $n_{id} =1$) in this situation only if $n$ is held approximately constant (for example by a contact, see supporting information Note 2).

| Recombination mechanism | Trap energy | Carrier overlap | $β$ | $γ$ | $n_{id}$ |
|---|---|---|---|---|---|
| band-to-band | N/A | $n >> p$ | 1 | 1 | 1 |
| band-to-band | N/A | $n = p$ | 2 | 2 | 1 |
| SRH | shallow | $n >> p$ | 1 | 1* | 1* |
| SRH | shallow | $n = p$ | 2 | 2 | 1 |
| SRH | mid-gap | $n >> p$ | 1 | 1 | 1 |
| SRH | mid-gap | $n = p$ | 2 | 1 | 2 |

A device that shows a change in ideality as a function of time following preconditioning, implies that there has either been a shift in the dominant recombination reaction order (for example from SRH to band-to-band dominated) or a change in the charge density dependence of the quasi Fermi-level splitting as defined by the parameter *β*.

In working solar cells, non-radiative recombination via defect states in the bandgap is consdered the most important recombination pathway since it can be mitigated via passivation of defects, particularly at grain boundaries and interfaces.[15,16] In *p-i-n* devices with a built-in electric field, the overlap in the populations of electrons and holes at the interfaces is likely to be small, hence idealities close to 1 are often correlated with surface recombination while idealities closer to 2 are identified with bulk recombination via traps (see Table 1). In organic solar cells, experimental values of between 1 and 2 have frequently been observed suggesting a range of behaviours from surface to bulk recombination.[17,18] Idealities of > 2 have previously been attributed to a tail of states in the bandgap[19] or non-linear shunt resistance,[20] although Kirchartz *et al*. have shown these processes are unnecessary to reproduce unusually high idealities in thin devices where the distribution of the carrier population overalp is highly dependent on the magnitude of the carrier densities.[21]

Given this framework for understanding the origins of ideality factors we now examine both measurements and simulations of time-dependent ideality factors in perovskite solar cells with multiple architectures and simulated recombination schemes. In particular, we focus on the influence of mobile ionic charge on localised electron and hole population overlaps, and how this alters the measured ideality factor with time.

**Methods**

*Devices*

Measurements were performed on devices with two different architectures:

1. FTO(400 nm)/compact-TiO$_2$(50 nm)/mp-Al$_2$O$_3$(250 nm)/CH$_3$NH$_3$PbI$_3$(200 nm)/ Spiro-OMeTAD(200 nm)/Au(100 nm) as described in reference [22] which we will refer to as mesoporous Al$_2$O$_3$ (mp-Al$_2$O$_3$).
2. FTO(400 nm)/compact-TiO$_2$(50 nm)/mp-TiO$_2$(250 nm)/CH$_3$NH$_3$PbI$_3$(200 nm)/ Spiro-OMeTAD(200 nm)/Au(100 nm) as described in reference [23] which we will refer to as mesoporous TiO$_2$ (mp-TiO$_2$).

This allows a comparison of recombination in a device containing a planar electron transporting material/perovskite interface with one in which the interface is likely to be extended.

*Suns-$V_{OC}$*

The history-dependent nature of the open circuit voltage in perovskite devices required an adaptation of the conventional Suns-$V_{OC}$ method for determining the ideality factor. Figure 1b shows the experimental timeline for the measurement. The TRAnsient and Charge Extraction Robot (TRACER) system, described previously,[24] was used to precondition the measured cell for a set time (typically 100 s) at a specified prebias voltage, $V_{pre}$, prior to the measurement of $V_{OC}$ at each light condition. The cell was then simultaneously switched (with microsecond precision) to open circuit and a continuous light intensity (with photon flux $\varphi$) generated by a bank of white LEDs was switched on. The evolution of $V_{OC}(t)$ was then recorded using a National Instruments USB-6251 data acquisition card where $t$ is the time after illumination. This procedure was repeated sequentially for each light intensity used in the measurement from low to high intensity. The transient ideality factor, $n_{id}(t)$, of the solar cell for a given precondition was evaluated from the fit to the slope of the $V_{OC}(t)$ against $\ln(\varphi)$ (equation 2) between $\varphi$ = 0.2 – 2 sun equivalents (calibrated with the photocurrent generated by a solar simulator) determined at different delay times $t$.

*J-EL measurement*

An alternative method to determine $n_{id}$ was also checked using electroluminescence *vs* current density measurements. Electroluminescence (EL) was measured using a Shamrock 303 spectrograph combined with an iDUS InGaAs array detector cooled to –90 °C. Before measurement of EL at each injection current density, the device was preconditioned at a voltage, $V_{pre}$, as described above for the Suns-$V_{OC}$ measurements. After this poling step, the device was set to a constant injection current level ranging between $J$ = 1.25 and 1250 mA cm$^{-2}$. A pair of optical lenses were used to focus the resulting EL signal allowing sampling at time intervals of between 0.02 – 0.1 s. The EL signals were normalised relative to the value at $t$ = 1 s. The relative emission flux for each injection current was subsequently determined by integration over the whole spectrum. The EL emission flux $\varphi_{EL}$ can be described as:

$$\phi_{EL} = \frac{J_{0,rad}}{q} \exp\left(\frac{qV}{k_B T}\right) \tag{7}$$

which is proportional to the radiative component of the total recombination. Here, $J_{0,rad}$ is the dark saturation current for the radiative recombination only. Since the radiative component of recombination comes from band-to-band transitions, equation 7 has an effective ideality factor of unity (c.f. Table 1). We can combine equation 7 with the diode equation 1 (where, with no illumination, $J_{SC}$ = 0) to obtain an expression for ideality factor $n_{id}$ of the device:

$$n_{id}(t) = \frac{d \ln \phi_{EL}(t)}{d \ln J} \tag{8}$$

The ideality factor can therefore be derived from the slope of the logarithm plot of the EL intensity against injection current density.

*Suns-PL measurement*

As a further cross-check for the determination of $n_{id}$ we used photoluminescence (PL) intensity as a function of excitation intensity measurements. PL from perovskite devices was measured with the same spectrograph and detector system as the EL setup described above. A 473 nm diode laser was used as the excitation source, and a set of neutral density filters were applied to adjust the laser intensity. Prior to each measurement, the devices were preconditioned at $V_{pre}$ in the dark (as described for the Suns-$V_{OC}$ and *EL-J* measurements). The sample was then illuminated and the integrated PL signal recorded with a sampling interval of 0.1 s. A small fraction of the total recombination is in the form of radiative emission, where PL emission flux from the solar cell is related to the quasi Fermi-level splitting ($qV_{OC}$) in the device such that:

$$\phi_{PL} = \frac{J_{0,rad}}{q} \exp\left(\frac{qV_{OC}}{k_B T}\right) \tag{9}$$

During the PL measurement at open circuit, the total recombination flux, $U$, in the device will be proportional to the illumination flux from the laser, $\varphi$, so that using equation 1 for open circuit conditions ($J$ = 0 and $J_{SC} = q\varphi$) we again arrive at equation 2. By combining equation 9 with equation 2 we can derive the following expression:

$$n_{id}(t) = \frac{d \ln \phi_{PL}(t)}{d \ln \phi} \tag{10}$$

The cell ideality can therefore be evaluated from the slope of the logarithm of PL intensity (flux) plotted against the logarithm of the laser intensity (flux) at different times (*t*) following illumination.

*Transient photovoltage measurements*

To probe the direction of charge transport within different regions of the device we used the TRansient Of The TRansient Photovoltage (TROTTR) technique previously introduced in reference[4], using a 488 nm laser excitation source superimposed on the white LED bias light. The measurement is similar to the Transient Suns-$V_{OC}$ measurement, performed at 1 sun equivalent intensity, with the superposition of short (0.5 - 10 µs) 488 nm laser pulses at periodic intervals (1 second) throughout the background illumination stage to induce small perturbation photovoltage transients Δ$V$. The experimental timeline is reproduced from reference [4] in supporting information, Figure S7 where the experimental setup is also given (Figure S8). The relatively short penetration depth of the laser results in charge carriers generated close to the electrode which is directly illuminated by the pulse.

A negative Δ$V$ indicates the presence of a reverse electric field in the region of the active layer probed by the excitation pulse.

*Simulation*

The DrIFtFUSION simulation tool was used to solve for electron, hole and mobile ion densities as a function of position and time in a *p-i-n* device structure. Here, a mirrored-cell approach with open circuit ($J = 0$) boundary conditions, was used to enable the open circuit voltage to be solved for directly as a function of time.[4] Prior to solving with illumination, the equilbrium solution for the cell in the dark with an applied voltage of $V_{pre}$ was established to give the initial condition for the subsequent $V_{OC}(t)$ solutions. For simplicity a uniform generation profile was used during the $V_{OC}$ simuations and solutions were caculated for 9 different constant light intensities ranging from 0.1 to 25.6 sun equivalents where (1 sun equivalent was set to be a generation rate $2.5 \times 10^{21}$ cm$^{-3}$s$^{-1}$ throughout the 400 nm thickness of the perovskite layer). A single positive mobile ionic defect species was considered with a uniform static distribution of negative counter-ions in the active layer. Low rates of second order band-to-band recombination ($U_{btb} \propto np$) were implemented in all layers, whereas SRH recombination ($U_{SRH}$) was used in different locations of the device dependent on the recombination scheme. Five different recombination schemes were examined summarised in Table 2. The parameteres used and a more complete description of the model are given in the supporting information Table S1 and reference [4]. The MATLAB code is freely available to be downloaded at https://github.com/barnesgroupICL/Driftfusion.

**Table 2.** Recombination mechanisms simulated in the study. Shallow traps were set to be in the bandgap and 0.2 eV away from the conduction and valence bands in the *n* and *p*-type layers respectively. Mid-gap traps were defined to be in the centre of the 1.6 eV bandgap.

| Scheme number | Dominant recombination mechanism | Trap energy |
|---|---|---|
| 1 | band-to-band only | N/A |
| 2 | interface SRH | shallow |
| 3 | interface SRH | mid-bandgap |
| 4 | bulk SRH | shallow |
| 5 | bulk SRH | mid-bandgap |

The recombination coefficients were adjusted dependent on the trap energy to yield $V_{OC}$ values approximately similar to those observed experimentally. Where trap levels are described as 'shallow' the energies were set to be in the bandgap, 0.2 eV away from the conduction and valence bands in the *n* and *p*-type layers respectively. While symmetrical trap distributions such as this are highly improbable in reality, this approach enabled the study of location-dependent recombination in isolation of asymmetric electron and hole recombination rates.

The simulated solutions for $V_{OC}(t)$ determined for the different photon fluxes were used evaluate $n_{id}(t)$ from the slope of $V_{OC}$ *vs* the logarithm of $\varphi$. Since this slope varied somewhat over the range of $\varphi$, particularly with low $V_{pre}$ values, the average value is quoted in this study (see supporting information, Figure S10).

**Results and Discussion**

*The evolution of $V_{OC}$ with light intensity and preconditioning bias*

Examples of the measured evolution of the $V_{OC}$ following illumination at different constant light intentisities are shown in Figure 2a and Figure 2b pre-biassing at 1.2 V and 0 V respectively in the dark for a mesoporous $Al_2O_3$ cell. It is apparent that, following $V_{pre}$ = 1.2 V, the $V_{OC}$ gradually decreases over the course of 60 s from an initially higher value towards a steady-state value for each light intensity. In contrast, following $V_{pre}$ = 0 V the $V_{OC}$ rises towards a steady-state value. Figure 2c shows that the slope of $V_{OC}$ plotted against the logarithm of photon flux density increases with time after illuination for the $V_{pre}$ = 1.2 V case.

We note that when the measurement protocol was applied to devices with poor degradation stability, we observed irreversible losses during the measurement, these typically resulted in variation in initial-value of the ideality (see supporting information Note 4). When the measurements were applied to more stable devices we observed repeatability in the transient ideality factor signature, and were able to correlate the results with our simulations.

We have previously discussed the mechanism underlying the evolution of $V_{OC}$ at 1 sun equivalent intensity following different preconditioning biases in considerable detail.[4] The gradual change in $V_{OC}$ often observed in perovskite cells can be explained by a slow redistribution of ionic charge from an initial state towards a new dynamic equilibrium at open circuit in the light. In the initial state the charge is accumulated at the interfaces, screening the internal potential resulting from the combination of the built-in and preconditioning voltages. Upon illumination, the change in the electric field distribution within the device, resulting from the development of a photovoltage, can result in either enhanced segregation of charges towards the charge collecting contacts (following positive $V_{pre}$ close to the open circuit potential) or segregation of photogenerated charges towards the opposite contacts (following zero or negative $V_{pre}$). If these ion distribution effects are combined with significant interfacial recombination, the $V_{OC}$ either slowly decreases or increases from its initial value as the mobile ionic charge moves to a new equilibrium distribution. At the new equilibrium the internal electric field within the perovskite layer is again screened by an accumulation of ionic charge at the interfaces and a new $V_{OC}$ value is reached. It is likely that this core process of ionic redistribution is present in all the devices examined in this work. The rate of redistribution will depend on the light intensity and magnitude of the preconditioning bias. The details of how this redistribution influences the evolution of the transient ideality factor will, in turn, depend on the details of the recombination processes in the device.

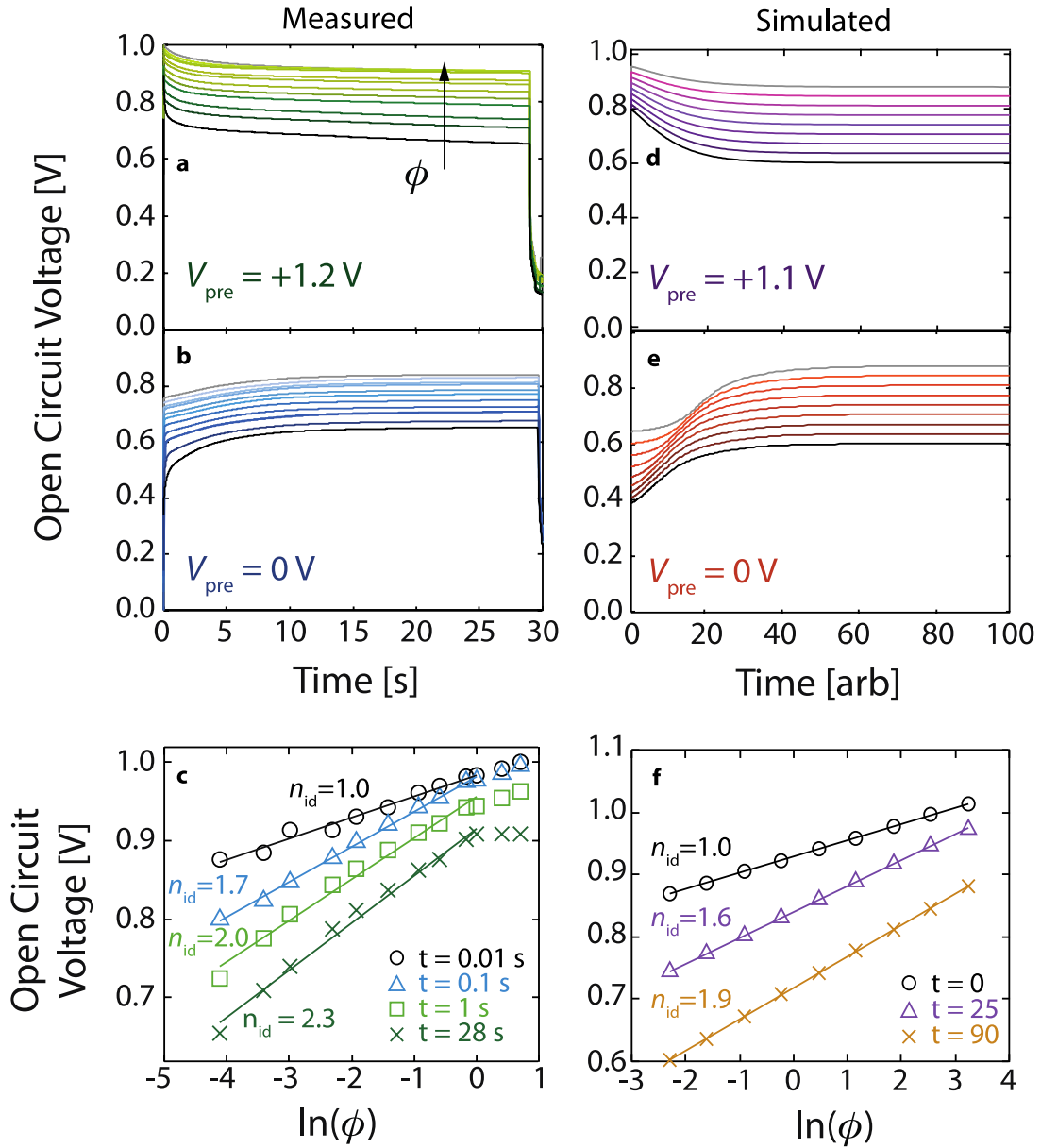

**Figure 2. Temporal evolution the open circuit voltage in measured and simulated devices.**
Examples of the evolution of measured $V_{OC}$ for different incident photon fluxes ($\varphi$ = 0.01 – 2 sun equivalents) with time following dark preconditioning with (**a**) $V_{pre}$ = 1.2 V and (**b**) $V_{pre}$ = 0 V for a mesoporous $Al_2O_3$ cell. (**c**) The measured $V_{OC}$ plotted against incident photon flux, $\varphi$, at different delay times following illumination for the $V_{pre}$ = 0 V case shown in (**a**). The ideality factor shows an increase in time. (**d**) The simulated evolution of the $V_{OC}$ for different incident photon fluxes ($\varphi$ = 0.1 – 25.6 sun equivalents) from a $V_{pre}$ = 1.1 V dark equilibrium solution and (**e**) from a $V_{pre}$ = 0 V dark equilibrium solution. For both preconditioning cases shown in the example, recombination scheme 3 was used in the simulation (Table 2, band-to-band in bulk and interfacial SRH recombination *via* mid-gap states). (**f**) Simulated $V_{OC}$ *vs* incident photon flux, $\varphi$, at different delay times following illumination for the dark $V_{pre}$ = 1.1 V initial conditions case shown in (**d**) also shows an increase in ideality factor with time.

*The evolution of the transient ideality factor*
Figure 3a shows the evolution of $n_{id}(t)$ for the two different types of cell following dark preconditioning at $V_{pre}$ = 1.2 V (mp-$Al_2O_3$ device), $V_{pre}$ = 1.0 V (mp-$TiO_2$ device). It is apparent that the evolution of $n_{id}(t)$ is significantly different for each of the examples shown. Following $V_{pre}$ = 1.2 V the ideality starts at $n_{id}(t=0) \approx 1$ and increases to approximately 2.1 for the mesoporous Al$_2$O$_3$ cell (green curve, circle markers) during the 30 seconds of the measurement. When the ideality for the same cell was evaluated following $V_{pre}$ = 0 V, the values of $n_{id}$ were much higher that expected given the classical interpretation, starting at 3.7 and decreasing to approximately 1.8 (see supporting information, Figure S9a). The mesoporous TiO$_2$ cell (light blue curve, triangle markers) also showed and increase in $n_{id}$ following $V_{pre}$ = 1.0 V, from 1.5 to around 2.7. The curve appears similar in shape to that of the mesoporous Al$_2$O$_3$ device but shifted with respect to the y-axis. Following $V_{pre}$ = 0 V, the $n_{id}$ again shows much higher values, starting at 3.2 and finishing at 2.6 (see supporting information, Figure S9a). A summary of the initial and final $n_{id}$ values for different measurement techniques and preconditioning voltages is given in Table 3.

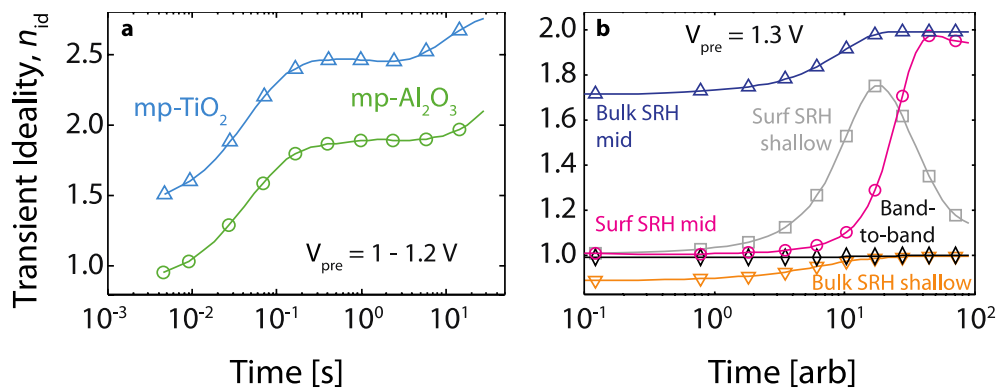

**Figure 3. Measured and simulated evolution of the ideality factor with time following forward biasing.** Measured $n_{id}(t)$ for mesoporous TiO$_2$ (light blue curve, triangle markers), mesoporous Al$_2$O$_3$ (green curve, circle markers) cells following dark (**a**) $V_{pre}$ = 1.2 V (mp- Al$_2$O$_3$ device), $V_{pre}$ = 1.0 V (mp-TiO$_2$ device). (**b**) The simulated transient ideality factor, $n_{id}(t)$, derived from simulations of $V_{OC}(t)$ for different incident photon fluxes in a *p-i-n* structure starting from initial conditions simulated for dark equilibrium with $V_{pre}$ = 1.3 V. The results from simulations with different dominant recombination mechanisms are shown: band to band recombination (black curve, diamond markers), interfacial SRH recombination *via* shallow traps (grey curve, square markers), interfacial SRH recombination *via* deep traps (pink curve, circle markers), bulk SRH recombination *via* shallow traps (yellow curve, inverted triangle markers), and bulk SRH recombination *via* deep traps (dark blue curve, triangle markers).

**Table 3. Summary of the instantaneous and steady state ideality factors measured for different devices using different techniques.**

| Cell Architecture | Precondition voltage, $V_{pre}$ (V) | Instantaneous ideality, $n_{id}(t≈0)$ | | | 1 sun equivalent *J-V* performance | |
|---|---|---|---|---|---|---|
| | | Suns-$V_{OC}$ | *J*-EL (0.02 s) | Suns-PL (0.01 s) | Apparent PCE (%) | $V_{OC}$ (V) |
| mp-AL$_2$O$_3$ | 1.1 | 1.0 | 0.9 | 0.9 | 10.5 | 1.03 |
| MAPIC | 0 | 3.6 | - | 2.0 | 6.7 | 0.94 |
| mp-TIO$_2$ | 1.1 | 1.5 | 1.7 | 1.6 | 9.7 | 0.86 |
| MAPIC | 0 | 3.2 | - | 1.8 | 7.1 | 0.81 |

*Comparison with instantaneous idealities determined from electroluminescence and photoluminescence measurements*

To cross-check that the values of $n_{id}$ measured this way are reasonable we made further measurements of the initial value of the ideality factor $n_{id}(t≈0)$ following dark prebiassing using both electroluminescence *vs* current density measurements (*J*-EL) and photoluminescence *vs* illumination intensity measurements (Suns-PL). Figure 4 shows examples of these measurements which are summarised for various cell types in Table 3. There was good agreement between the ideality values determined using the different methods following $V_{pre}$ > 0. For $V_{pre}$ = 0, we noted discrepancies between the Suns-PL measurements and Suns-$V_{OC}$ values in some cases leading us concentrate our analysis on transient ideality factors derived for $V_{pre}$ > 0 (this is discussed further along with possible explanations in the supporting information Note 2). Overall, the agreement between the different techniques suggests the concept of a transient ideality factor measured by the Suns-$V_{OC}$ technique is related to recombination processes and not to charge storage or capacitive effects within the cell.

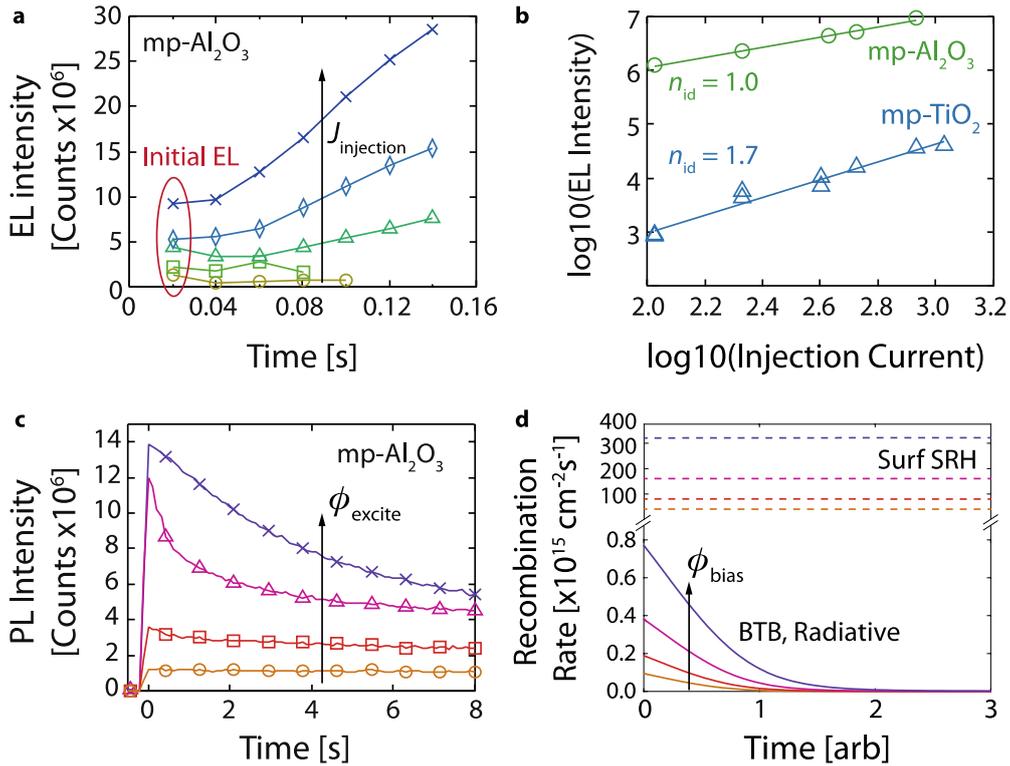

**Figure 4. Alternative measurements of the initial transient ideality factor using electroluminescence and photoluminescence.** (**a**) Integrated electroluminescent emission flux density *vs* time for different injection current densities, $J$ = 107, 213, 426, 533 and 853 mA cm$^{-2}$ following the application of $V_{pre}$ = 1.1 V to a mesoporous Al$_2$O$_3$ cell. (**b**) Evaluation of $n_{id}$ for $t$ = 0 – 0.02 s from EL emission flux density *vs* injection current density for the three architectures of device. (**c**) Integrated photoluminescent flux density *vs* time for different excitation flux densities following the application of $V_{pre}$ = 1.2 V to a mesoporous Al$_2$O$_3$ cell. (**d**) Simulated integrated band-to-band (BTB, radiative, solid curves) and interfacial SRH recombination (dashed curves) fluxes as a function of time for a cell with recombination scheme 3 for $V_{pre}$ = 1.3 V.

*Comparison with simulated transient ideality factors*

Different recombination mechanisms might be expected to dominate different device architectures. If the transient ideality factors (plotted in Figure 3a and Figure 3b) are related to these mechanisms then we hypothesise that the shape of $n_{id}(t)$ gives a signature which can be used to identify the primary recombination pathway for any given device. To test this possibility we now examine simulations of the $n_{id}(t)$ measurement for modelled cells with a range of recombination schemes as summarised in Table 2. In particular, the simulations provide us with a means to assess the likely consequences of mobile ionic defects on our measurements. Furthermore, by examining the evolution of the free electron and hole profiles as the mobile ionic charge redistributes, we can ascertain how the relative concentration of electrons and holes varies with time in regions of the cell where recombination is localised. This allows us to diagnose the specific conditions in Table 1 that give rise to an observed value of $n_{id}$ at a given time in the measurement.

Figure 2d and Figure 2e show examples of the evolution of $V_{OC}$ as a function of time after illumination for different light intensities for two initial conditions corresponding to $V_{pre}$ = 1.1 V and 0 V for a simulated device dominated by interfacial SRH recombination *via* mid-gap states (scheme 3 in Table 2). The results reproduce the majority of features seen in Figure 2d and Figure 2e, and can be used to find $n_{id}$ following different illumination times (Figure 2f). This process was repeated for the different recombination schemes shown in Table 2 allowing $n_{id}(t)$ to be evaluated from the results for the two initial conditions. The corresponding simulated $n_{id}(t)$ curves are shown in Figure 3b. We note that linearity of the $V_{OC}$ vs $\ln(\varphi)$ curves used to determine $n_{id}(t)$ is considerably greater following the forward bias precondition relative to the zero bias condition (see examples in the supporting information, Figure S10). Combined with the descrepancy between the ideality following $V_{pre}$ = 0 V for different measurement techniques discussed in supporting information, Note 3 this suggests that $n_{id}(t)$ determined following forward biasing provides a more reliable measure of recombination mechanism.

There are a number of strikingly similar features between particular pairs measured and simulated $n_{id}(t)$ profiles in Figure 3. On the basis of these similarities and the more detailed analysis of the simulation output, discussed below, we assign the evolution of the measured $n_{id}(t)$ profiles examined in our study to likely dominant recombination schemes. We do not assert here that the dominant recombination mechanisms we assign to the devices will always represent the dominant recombination route for the respective architectures since it is very likely that variation in the processing during device fabrication can lead to a number of possible recombination routes and thus device-to-device variation in $n_{id}(t)$ for cells prepared with the same architecture (discussed further in the supporting information, Note 4). Rather, the approach described here presents a means to assess the most likely recombination routes for a particular device.

*Recombination mechanism in the mesoporous $Al_2O_3$ cell*
The transient ideality signatures for the mp-$Al_2O_3$ cell (Figure 3, green curve, circle markers) correspond most closely to a model where interfacial SRH recombination *via* mid-bandgap states (scheme 3) is the dominant recombination mechanism. This recombination scheme is also consistent with the large degree of *J-V* hysteresis seen in this device (see supporting information, Figure S13a) since we, and others, have previously shown that a combination of mobile ionic charge and interfacial recombination are necessary to observe hysteresis.[4,7,8,12,25] While computational studies on defective $CH_3NH_3PbI_3$ crystals have not shown deep trapping states with easily accessible activation energies,[26,27] the interface between $CH_3NH_3PbI_3$ and the compact $TiO_2$ layer is likely to contain a high density of deep inter-bandgap electronic states as commonly observed at the $TiO_2$ hole transporting material interface in dye sensitised solar cells.[28,29] Furthermore, recent studies have shown that recombination rates at the hole transport layer interface are significantly influenced by doping[30] and that the $CH_3NH_3PbI_3$ interface with heavily lithium-doped Spiro-OMeTAD is a site of significant recombination.[7]

We now consider the likely mechanism underlying the observed changes in the transient ideality factor over the course of the measurement for this device. In the simulation with recombination scheme 3, following $V_{pre}$ = 1.3 V, $n_{id}(t)$ initially starts at a value close 1.0 and increases to a value of close to 2 by the end of the measurement (Figure 3b, green curve, circle markers). The evolution of the simulated integrated recombination fluxes in the cell indicate that the SRH recombination at the contacts dominates at all times during the transient (see Figure 4d) such that the changes in the ideality factor cannot be explained by a change in the dominant recombination mechanism. We note however that there is significant net reduction in the fraction of recombination arising from band-to-band recombination during the simulation, which is consistent with the reduction in PL signal with illumination time observed over a timescale of seconds in the mp-Al$_2$O$_3$ device (Figure 4c and Figure 4d).

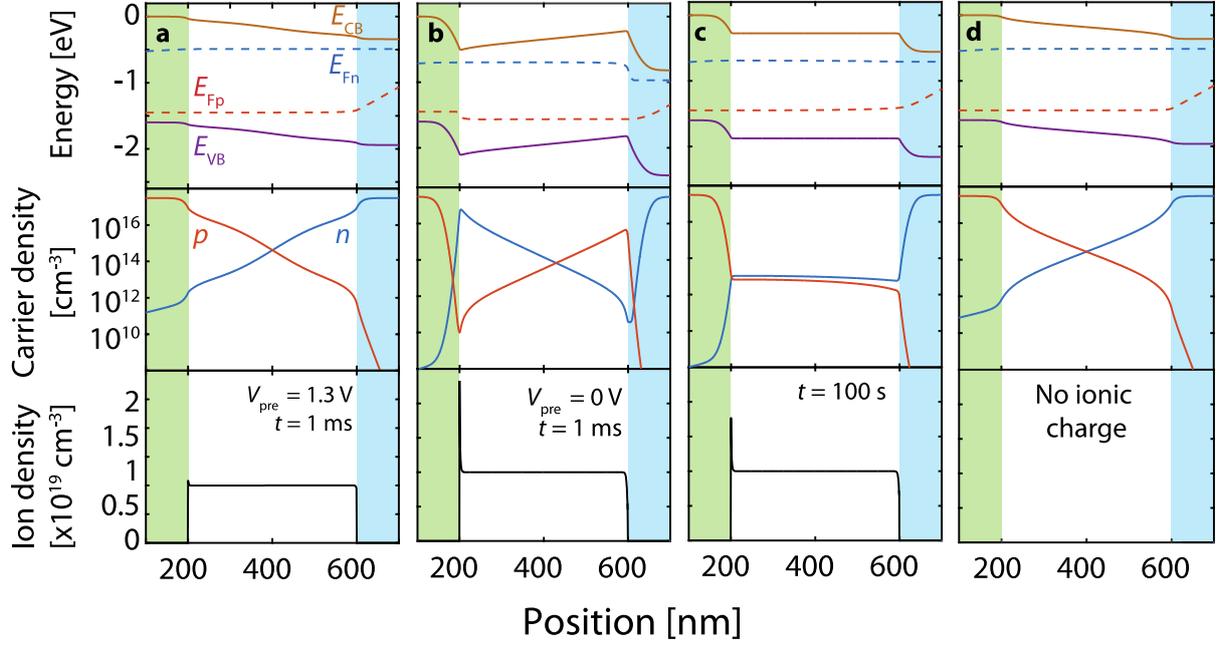

**Figure 5. Example energy level diagrams and charge density profiles during the open circuit voltage transient simulation.** Energy level diagrams, electron ($n$), hole ($p$) and mobile ion densities after simultaneously illuminating and switching to open circuit with $\varphi$ = 1.6 sun equivalent for a device preconditioned with (**a**) $V_{pre}$ = 1.3 V in the dark following $t$ = 1 ms of illumination at open circuit. A device preconditioned with $V_{pre}$ = 0 V in the dark following (**b**) $t$ = 1 ms and (**c**) $t$ = 100 s of illumination at open circuit. After 100 s of illumination the ions have migrated from their preconditioned distribution to establish a new quasi-equilibrium such that the profiles of $V_{pre}$ = 1.3 V after $t$ = 100 s are identical to those shown in (**c**). (**d**) A device with the same parameter set without mobile ions. Recombination scheme 3 (Table 2) was used in the simulation results shown, the green and blue shaded regions indicate the respective $p$ and $n$-type regions of the device.

Since the primary recombination mechanism does not change during the measurement (and thus $\gamma$ is constant) the observed variation in $n_{id}$ is likely to be related to variation in $\beta$ (c.f. equation 7), which defines the relationship between the charge carrier density and the perturbation of the quasi Fermi-levels from equilibrium. Examination of the initial (Figure 5a) and final (Figure 5c) simulated electron and hole profiles for the $V_{pre}$ = 1.3 V case reveals that there is indeed a change in the relative electron concentrations in the interface regions as the ions redistribute. Initially $n \gg p$ in at the $n$-type interface (and $p \gg n$ at the p-type interface) such that for SRH interfacial recombination *via* mid-bandgap traps $\beta \approx 1$ (see Table 1). In the final state $n \approx p$ at both interfaces such that $\beta \approx 2$, explaining the change from 1 to 2 in $n_{id}(t)$ during the course of the measurement.

The similarity in electronic charge density profiles following preconditioning at $V_{pre}$ = 1.3 V (Figure 5a) and that of a simulated device without mobile ionic charge (Figure 5d) is highly significant; the preconditioning stage reduces the density of ionic charge accumulated at the interface with respect to $V_{pre}$ = 0 V (Figure 5b) and results in an electric field in the device at $t$ = 0 s of the measurement which drives charge carriers towards their respective transport layers. Consequently, forward biasing with an applied potential sufficient to separate the electron and hole densities at the interfaces determines an instantaneous ideality factor which can be interpreted using the standard theory (as outlined in the introductory theory section). The steady-state $n_{id}$ value can only be interpreted correctly by considering the fact that $n \approx p$ ($\beta \approx 2$) throughout the active layer (see Table 1). The evolution of the transient ideality factor following $V_{pre}$ = 0 V is discussed further in the supporting information Figure S8 caption.

*Recombination mechanism in the mesoporous $TiO_2$ cell*
The transient ideality factor profile of the mp-$TiO_2$ (Figure 3a, light blue curve, triangle markers) following $V_{pre}$ = 1 V shows an intermediate behaviour between surface (Figure 3b, pink curve, circle markers) and bulk SRH recombination (Figure 3b, dark blue curve, triangle markers) *via* mid-bandgap traps within the bulk of the active layer.

The shape of the simulated SRH mid-gap recombination in the bulk $n_{id}(t)$ signatures can be understood by examining the charge carrier concentration profiles during the simulations. These were qualitatively similar to those shown in Figure 5 both for the $V_{pre}$ = 1.3 V and $V_{pre}$ = 0 V cases. For the measurement following forward bias preconditioning, the initial recombination flux has significant contributions from regions of the bulk where $n \neq p$ such that $1 < n_{id} < 2$. In the final state $n \approx p$ throughout the bulk such that $n_{id} \rightarrow 2$. The simulated measurement following $V_{pre}$ = 0 V in this case resulted in $V_{OC}$ *vs* $\ln(\varphi)$ slopes that were very non-linear so the relevance of the derived $n_{id}(t)$ values to recombination values are questionable. We observe a curious case where the initial recombination is dominated by the central region where $n \approx p$ (so that $\gamma \approx 1$) but Fermi level splitting is dominated by changes in concentration near the interfaces ($\beta \approx 1$) so that the mean ideality goes from 1 to 2 as the $\beta \rightarrow 2$ during the redistribution of ionic charge to a steady state (see supporting information, Figure S11).

*Determining the spatial extent of the charge collection interface/recombination region*
In order to further investigate the origin of the initial measured ideality value of $n_{id}(t = 0)$ = 1.5, we probed the internal electric field by performing Transient of the Transient Photovoltage measurements (see Methods) on similar mp-$TiO_2$ and mp-$Al_2O_3$ architecture devices fabricated using a Transparent Conductive Adhesive (TCA) top electrode[31] with a short wavelength laser excitation source. The absorption coefficient for $CH_3NH_3PbI_3$ at 488 nm is 1.84 x $10^5$ $cm^{-1}$, giving a characteristic penetration depth of 54 nm (see optical modelling in supporting information Figure S14). Thus, only charge carriers generated close to the contact through which the device is illuminated are expected contribute to change in photovoltage ($\Delta V$).

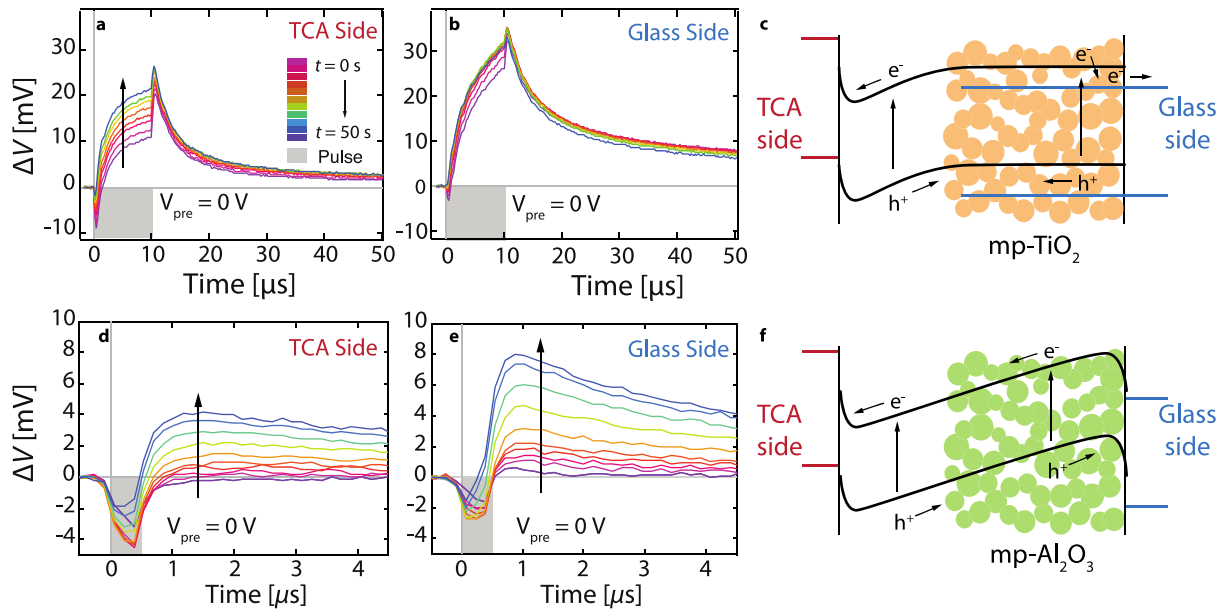

**Figure 6. Transient photovoltage measurements on bifacial devices and band energy level schematics for mp-TiO$_2$ and mp-Al$_2$O$_3$ architectures.** Transient photovoltage Δ$V$ measurements for following $V_{pre}$ = 0 V (see experimental timeline in **supporting information, Figure S7**). The mp-TiO$_2$ device shows an initial negative deflection to Δ$V$ when excited through (**a**) the transparent conductive adhesive (TCA) side by the 488 nm laser pulse. (**b**) The change in signal is much smaller when the device is excited through the glass side, implying an inverted electric field is localised to the TCA side as shown in (**c**) the band schematic. The mp-Al$_2$O$_3$ device showed similarly negative initial Δ$V$ responses when illuminated through both (**d**) TCA and (**e**) glass sides, implying the inverted field is extended into the electronically insulating mp-Al$_2$O$_3$ region as shown in (**f**) the device schematic. mp-TiO$_2$ and mp-Al$_2$O$_3$ are indicated by the yellow and green shaded areas respectively. Spiro-OMeTAD, perovskite, and TiO$_2$ conduction and bands are shown as red, black and blue curves respectively (energies not to scale).

Figure 6 shows the results of the TPV experiments combined with schematics of the likely band configuration on the mp-TiO$_2$ and mp-Al$_2$O$_3$ devices. In both cases the cells were preconditioned at $V_{pre}$ = 0 V for 60 s prior to illumination at 1 sun equivalent intensity. The configuration of ions under these conditions would be expected to initially induce a reverse field in the devices similar to that seen in the simulated device following $V_{pre}$ = 0 V (Figure 5b).[9,10] At early times (<10 s) following illumination through the TCA side of the device, the TPV signal for the mp-TiO$_2$ showed a distinct negative deflection (Figure 6a). This negative response was significantly less pronounced when the device was instead illuminated through the glass/FTO side (Figure 6b). We interpret this as evidence that the electric field is predominantly confined to the pure perovskite phase in the device containing mp-TiO$_2$ as shown in Figure 6c. Here, carriers generated close to the hole transporter are driven in the wrong direction with respect to charge collection, leading to an increased accumulation of minority charge carriers at the Spiro-OMeTAD/Perovskite interface, increased recombination, and an associated negative TPV deflection and reduced magnitude as compared to later times in the measurement. Carriers generated in the mesoporous region instead diffuse with towards the correct contact and generate a positive signal. After 50 s the majority of the mobile ionic charge has migrated to screen out the field once more and the signal becomes similarly positive independent of illumination side (Figure 6a and Figure 6b, blue curves).

When the same experiment was repeated on the mp-Al$_2$O$_3$ device, a negative deflection was apparent in the initial TPV signals independent of illumination side (Figure 6d and Figure 6e), suggesting that the field is extended throughout the perovskite in the mesoporous region (Figure 6f). During illumination under open circuit conditions, the TPV signal followed a similar evolution to a more positive signal independent of illumination side in the proceeding 50 s. The persistent negative signal during the pulse after 50 s implies a long-lived polarisation in the device that could result from slower moving ionic species.[32]

Following forward bias preconditioning at $V_{pre}$ = 1 V, the negative component of the signal was diminished in all cases indicating that the strength of the inverted field at open circuit had been reduced (see supporting information, Figure S15), consistent with the simulation following forward bias preconditioning in Figure 5a.

Based on these results we hypothesis that in the mp-TiO$_2$ device, at the Spiro-OMeTAD interface the charge populations are separated ($β ≈ 1$), while within the mesoporous region and at the c-TiO$_2$ interface, the charge populations are approximately equal ($β ≈ 2$) following a forward bias precondition. The scaffold is likely to introduce deep trapping states in the region of this extended interface, leading to a reaction order of $γ ≈ 1$. If both interfaces have similar SRH time constants then an ideality of $1 < n_{id} < 2$ is expected: While the device is still dominated by surface recombination, the overlapping electron and hole populations ($n ≈ p$) throughout the mesoporous region leads to an ideality factor greater than 1. Further work incorporating the scaffold region into models will be required to test the above hypotheses. The reduced hysteresis in the cell's *J*-*V* curve by comparison with the mp-Al$_2$O$_3$ device (see supporting information, Figure S13a and Figure S13c) also suggests that recombination is dominated by a mix of surface and bulk effects.[8]

To summarise, these observations imply that the extended interface introduced by the mesoporous TiO$_2$ allows for recombination of electrons and holes *via* deep traps throughout the mesoporous region which can be considered 'bulk like'. In contrast, cells with mesoporous Al$_2$O$_3$ appear to be dominated by deep trap recombination primarily confined to the charge collecting interfaces, and not extending through the insulating mesoporous layer.

**Conclusions**

We have introduced a method to evaluate the transient ideality factor of perovskite solar cells following a voltage preconditioning period in the dark. The approach allows meaningful information about the dominant recombination mechanism to be derived in the presence of mobile ionic defects. The mobile ions result in an evolving electrostatic profile throughout the thickness of a perovskite solar cell and consequently a time varying $V_{OC}$ which hampers measurement of an ideality factor by conventional means. By preconditioning cells at a forward bias potential sufficient to separate the charge carrier densities at the interfaces, the cell can temporarily be set in a state which would yield a similar ideality factor to a similar device which contained no mobile ions and could thus be analysed using the convention theories applied to ideality factors.

With the aid of time-dependent device simulations, which include the effects of mobile ions, we have shown that the evolution of the ideality factor following different voltage preconditions can be used as a signature to identify the dominant recombination mechanism in a perovskite solar cell. We demonstrated this idea with standard perovskite device architectures incorporating mesoporous $TiO_2$ and mesoporous $Al_2O_3$ scaffolds. For these examples we assigned surface recombination via deep traps at the charge collection layer interfaces as the dominant recombination mechanism in both device types. Transient photovoltage measurements on devices with dual transparent electrodes indicated that, distinct from the mp-$Al_2O_3$ device, the electric field was predominantly confined to the pure perovskite region in the mp-$TiO_2$ architecture. The homogeneity of electron and hole populations throughout the extended mesoporous $TiO_2$ interface resulted in a measured initial ideality following forward bias of 1.5 despite surface recombination remaining the dominant mechanism in the device.

We emphasise that the dominant recombination mechanism can vary between devices which were nominally prepared in the same way, an observation we attribute to small differences in the processing conditions and microstructure of the cells. We also note that our observations do not preclude the possibility that trap energy and spatial distribution in the device may be dependent on the distribution of ionic defects. Together our findings provide new insight on the meaningful interpretation of ideality measurements in perovskite solar cells and a valuable new tool for assigning the dominant recombination mechanism in devices.

**Acknowledgements**
We thank the EPSRC for funding this work (EP/J002305/1, EP/M025020/1, EP/M014797/1, EP/L016702/1).

**Supporting information**

**Note 1: Derivation of ideality factor, $n_{id}$ as a function of the charge carrier density relationship to Fermi level splitting, $β$ and the reaction order, $γ$**

Using the Shockley diode equation, the ideality factor, $n_{id}$ can be expressed in terms of the quasi-Fermi level splitting $ΔE_F$ (identical to $qV_{OC}$ in the zero dimensional model) and recombination rate, $U$ such that:

$$n_{id} = \frac{1}{k_B T} \frac{dΔE_F}{d \ln U}$$

where $q$ is the elementary charge, $k_B$ is Boltzmann's constant and $T$ is the temperature.

Substituting $ΔE_F = k_B T \ln\left(\frac{n^β}{n_i^2}\right)$ and $U = kn^γ$ yields:

$$n_{id} = \frac{1}{k_B T} \frac{d\left(βk_B T\left(\ln n - \ln n_i^{2-β}\right)\right)}{d\left(γ(\ln n + \ln k^{1-γ})\right)}$$

$$= \frac{β}{γ} \frac{d\left(\ln n - \ln n_i^{2-β}\right)}{d(\ln n + \ln k^{1-γ})}$$

Let:

$x = \ln n$
$b = \ln n_i^{2-β}$
$c = \ln k^{1-γ}$

$$n_{id} = \frac{β}{γ} \frac{d(x-b)}{d(x+c)}$$

Using the variable substitution $x = x' - c$,

$$n_{id} = \frac{β}{γ} \frac{d(x'-b-c)}{dx'} = \frac{β}{γ}$$

**Note 2: Derivation of reaction orders**

Here we examine how the reaction order $γ$ for the general form of the equation describing the recombination rate $U$ in terms of the electron concentration, $n$, and the reaction order dependent recombination rate coefficient:

$$U = kn^γ$$

depends on the recombination mechanism and relative concentration of electrons and holes as listed in table 1 of the main text.

*Band-to-band recombination*

For *n* = *p* the recombination rate, *U*, is given by:

$$U = k_{\text{btb}} np = k_{\text{btb}} n^2$$

where $k_{\text{btb}}$ is the band-to-band recombination rate constant hence *γ* = 2.

For *n* >> *p*, the majority carrier density can be considered as effectively constant, leading to:

$$U \propto p$$

hence *γ* = 1.

*Shockley Read Hall recombination*

The compact form of the SRH recombination expression is:

$$U = \frac{np - n_i^2}{\tau_n(p + p_t) + \tau_p(n + n_t)}$$

$\tau_n$ and $\tau_n$ are the trap electron and hole recombination lifetimes and $n_t$ and $p_t$ are the electron and hole densities when the Fermi level is at the trap level, $E_t$, for example for electrons:

$$n_t = n_i \exp\left(\frac{E_t - E_{\text{CB}}}{k_B T}\right)$$

where $E_{\text{CB}}$ is the conduction band energy.

*Shallow traps, n = p*

For shallow traps close to the conduction band, in most cases $E_{\text{Fn}} < E_t$, and $n \ll n_t$, $p \gg p_t$, and $p \ll n_t$. Where *n* = *p* the rate of recombination is determined by both charge carrier densities:

$$U \approx \frac{np}{\tau_n p + \tau_p n} \approx \frac{n^2}{\tau_p n_t}$$

hence *γ* = 2.

*Mid-bandgap traps, n = p*

For deep traps where, in the presence of a bias $E_{\text{Fn}} > E_t$ and as a result $n \gg n_t$ and $p \gg p_t$. If *n* = *p* then:

$$U \approx \frac{np}{\tau_n p + \tau_p n} = \frac{n}{(\tau_n + \tau_p)}$$

hence *γ* = 1.

*Shallow traps, n >> p*

For shallow traps close to the conduction band, in most cases $E_{Fn} < E_t$, and $n \ll n_t$, $p \gg p_t$, and $p \ll n_t$. If $n \gg p$ and $n$ is approximately constant (for example if $n$ is pinned by the contact concentration $n_{contact}$) then recombination is limited by the availability of holes:

$$U \approx \frac{np}{\tau_n p + \tau_p n_t} \approx \frac{n_{contact} p}{\tau_p n_t}$$

hence *γ* = 1. However this does not necessarily hold generally. In the cases considered in the main paper this implies that the above approximation will be valid following forward bias preconditioning, but not zero or reverse bias preconditioning.

*Mid-bandgap traps, n >> p*

For deep traps in the presence of a bias where $E_{Fn} > E_t$, then $n \gg n_t$ and $p \gg p_t$. In a situation where $n \gg p$, recombination is limited by the availability of holes:

$$U \approx \frac{np}{\tau_n p + \tau_p n} \approx \frac{p}{\tau_p}$$

hence *γ* = 1.

**Device simulation parameter sets**

| Parameter name | Symbol | Value | Unit | Ref. |
|---|---|---|---|---|
| Band gap | $E_g$ | 1.6 | eV | 33 |
| Built in voltage | $V_{bi}$ | 1.3 | V | 34 |
| Dielectric constant | $\varepsilon_s$ | 20 | | 35 |
| Mobile ionic defect density | $N_{ion}$ | $10^{19}$ | $cm^{-3}$ | 36 |
| Ion mobility | $\mu_a$ | $10^{-12}$ | $cm^2 V^{-1} s^{-1}$ | |
| Electron mobility | $\mu_e$ | 20 | $cm^2 V^{-1} s^{-1}$ | 37 |
| Hole mobility | $\mu_h$ | 20 | $cm^2 V^{-1} s^{-1}$ | 37 |
| p-type donor density | $N_A$ | $3.0 \times 10^{17}$ | $cm^{-3}$ | |
| n-type acceptor density | $N_D$ | $3.0 \times 10^{17}$ | $cm^{-3}$ | |
| Effective density of states | $N_0$ | $10^{20}$ | $cm^{-3}$ | 8 |

Table S1 Key simulation device parameters

**Recombination scheme parameters**

Trap energies are referenced to the electron affinity of the material

*Band-to-band recombination*

| Parameter | *p*-type | Intrinsic | *n*-type | Unit |
|---|---|---|---|---|
| $k_{btb}$ | $10^{-10}$ | $10^{-10}$ | $10^{-10}$ | $cm^3 s^{-1}$ |
| $\tau_n$ | - | - | - | s |
| $\tau_p$ | - | - | - | s |
| $E_t$ | - | - | - | eV |

*Surface SRH, shallow traps*

| Parameter | *p*-type | Intrinsic | *n*-type | Unit |
|---|---|---|---|---|
| $k_{btb}$ | $10^{-10}$ | $10^{-10}$ | $10^{-10}$ | $cm^3 s^{-1}$ |
| $\tau_n$ | $10^{-15}$ | - | $10^{-15}$ | s |
| $\tau_p$ | $10^{-15}$ | - | $10^{-15}$ | s |
| $E_t$ | -1.4 | - | -0.2 | eV |

*Surface SRH, mid-gap traps*

| Parameter | *p*-type | Intrinsic | *n*-type | Unit |
|---|---|---|---|---|
| $k_{btb}$ | $10^{-10}$ | $10^{-10}$ | $10^{-10}$ | $cm^3 s^{-1}$ |
| $\tau_n$ | $10^{-10}$ | - | $10^{-12}$ | s |
| $\tau_p$ | $10^{-10}$ | - | $10^{-12}$ | s |
| $E_t$ | -0.8 | - | -0.8 | eV |

*Bulk SRH, shallow traps*

| Parameter | *p*-type | Intrinsic | *n*-type | Unit |
|---|---|---|---|---|
| $k_{btb}$ | $10^{-10}$ | $10^{-10}$ | $10^{-10}$ | cm$^3$s$^{-1}$ |
| $\tau_n$ | - | $10^{-15}$ | - | s |
| $\tau_p$ | - | $10^{-15}$ | - | s |
| $E_t$ | - | -0.2 | - | eV |

*Bulk SRH, mid-gap traps*

| Parameter | *p*-type | Intrinsic | *n*-type | Unit |
|---|---|---|---|---|
| $k_{btb}$ | $10^{-10}$ | $10^{-10}$ | $10^{-10}$ | cm$^3$s$^{-1}$ |
| $\tau_n$ | - | $10^{-10}$ | - | s |
| $\tau_p$ | - | $10^{-10}$ | - | s |
| $E_t$ | - | -0.8 | - | eV |

While the nominal band-to-band rate coefficients were set reasonably high, it can be shown that SRH recombination would still dominate in all cases provided that $\tau_{SRH} \ll k_{btb}n$. We start by expressing recombination *U* as a sum of SRH and band-to-band recombination:

$$U = \frac{1}{\tau_{SRH}}n + k_{btb}n^2$$

If SRH dominates:

$$\frac{1}{\tau_{SRH}}n \gg k_{btb}n^2$$

This leads to the condition:

$$n \ll \frac{1}{\tau_{SRH}k_{btb}}$$

Hence where $\tau_{SRH} = k_{btb} = 10^{-10}$ as in Scheme 5, provided *n* < $10^{20}$ cm$^{-3}$, which is guaranteed by the choice of $10^{20}$ cm$^{-3}$ for the effective density of states, SRH will dominate.

**Note 3: Discrepancies between $n_{id}(t≈0)$ for $V_{pre}$ = 0 derived from Suns-$V_{OC}$ and Suns-PL measurements**

For the mp-TiO$_2$ cell we note that although the changes in $n_{id}$ values from Suns-PL and Suns-$V_{OC}$ were consistent as $V_{pre}$ was varied the absolute values of $n_{id}$ for $V_{pre}$ = 0 were lower with the Suns-PL measurements ($n_{id}(t≈0)$ = 1.8) than the corresponding Suns-$V_{OC}$ measurements ($n_{id}(t≈0)$ = 3.2). This difference in absolute values may be related to the relatively localised generation of charge carriers by the laser spot during the Suns-PL measurement which might avoid localised recombination shunts (dependent on the ion distribution). These shunts would not be avoided during the Suns-$V_{OC}$ measurements since charge carriers are generated over the whole cell area.

**Note 4: Device-to-device variability in the transient ideality factor**

A significant issue for perovskite solar cells is variation in the properties of devices nominally prepared by the same methods. This issue has been noted by many authors including Pockett *et al.* who observed significant differences in the steady-state ideality factor observed for devices prepared in an identical way.[5] We also observed a similar phenomenon, with differences in the initial ideality factor values as well as at steady-state. For example, for a batch of 6 planar compact TiO$_2$ cells manufactured in a different laboratory to those described above, initial idealities after forward biasing at $V_{pre}$ = 1.1 V were 0.97 < $n_{id}(t≈0)$ < 4.22 ($\bar{n}_{id}(t \approx 0)$ = 1.78, $\sigma_{nid}$ = 1.16), while for mp-TiO$_2$ architectures following the same preconditioning, the values were 0.04 < $n_{id}(t≈0)$ < 1.12 ($\bar{n}_{id}(t \approx 0)$ = 0.75, $\sigma_{nid}$ = 0.62, for 6 devices). Mesoporous TiO$_2$ devices with triple cation active layers showed greater consistency with initial idealities after forward biasing of 0.40 < $n_{id}(t≈0)$ < 1.04 ($\bar{n}_{id}(t \approx 0)$ = 0.77, $\sigma_{nid}$ = 0.32, for 3 devices). The values close to 1 here might suggest that recombination at the compact TiO$_2$ layer interface is more significant than at the mp-TiO$_2$/CH$_3$NH$_3$PbI$_3$ interface in these devices. However, the unexpectedly high $n_{id}$ and standard deviation suggest that the recombination kinetics at any given interface type are highly dependent on localised details of the interface formation during device fabrication. Degradation during measurements could account for particularly low ideality factors since the $V_{OC}$ transients were measured in order of intensity from lowest to highest. If degradation was continuous throughout the measurement $V_{OC}$s would be lower than expected at higher light intensity, reducing the measured $n_{id}$. Further work on high performing, stable devices will be required to more confidently ascertain whether or not the signature transient idealities presented in this study allow more general statements on identifying the dominant recombination mechanism in different perovskite device architectures and material combinations.

**Transient of the Transient Photovoltage Experimental Timeline**

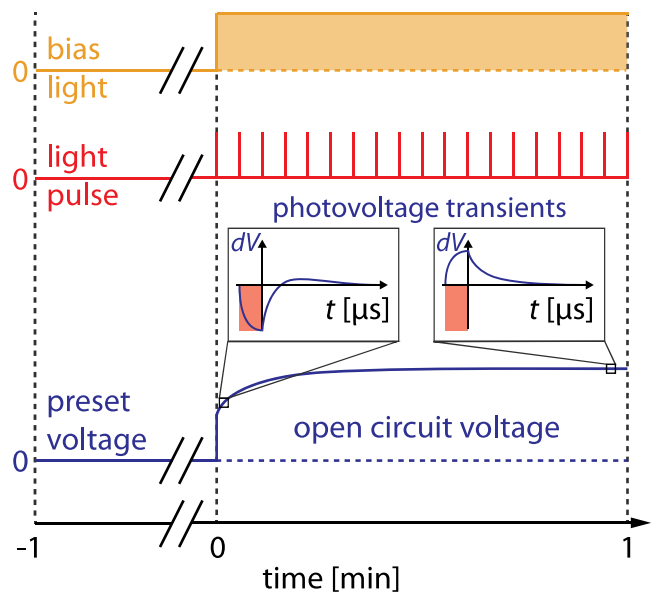

**Figure S7. Transients of the Transient photovoltage (TROTTR) measurement.** The device is held in the dark (the bias light state is represented by the yellow line and shaded regions) at a preset voltage, $V_{pre}$ (dark blue line) for 1 min before being switched to open circuit with 1 sun equivalent illumination. During the $V_{OC}$ evolution the cell is pulsed with a laser (red line and pink shaded regions) at 1 second intervals and the resulting photovoltage transients acquired.

**Transients of the Transients Experimental Setup**

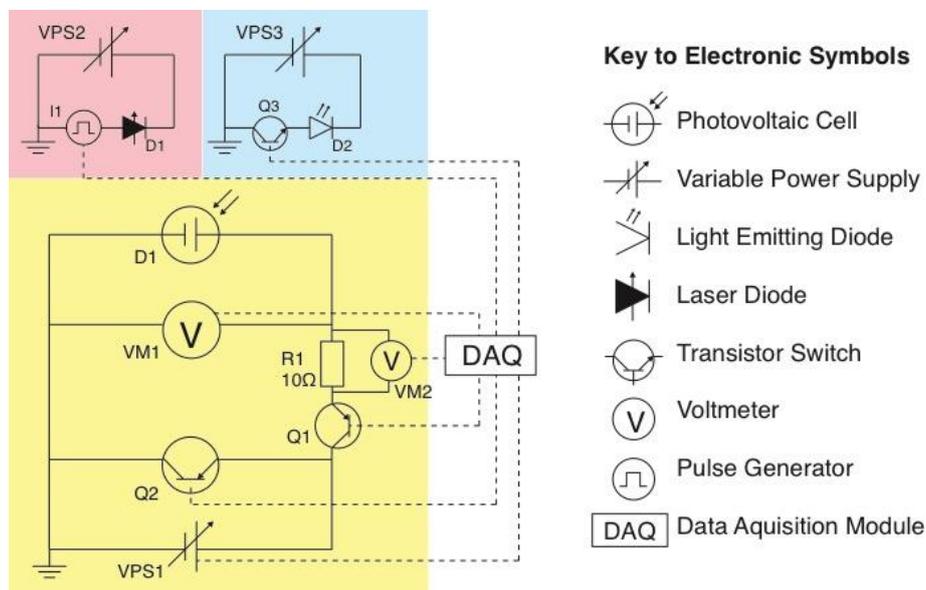

**Figure S8. Transient optoelectronic measurement system circuit diagram.** Circuit diagram for the transient data acquisition system. The yellow shaded area is the primary circuit, which allows the cell to be electrically biased and switched between open and closed circuit. The light red shaded area is the laser circuit and the light blue shaded region is the white bias LED circuit.

**Note 5: Transient system technical specifications**

Data acquisition was performed by a Tektronix DPO5104B digital oscilloscope and a National Instruments USB-6361 DAQ. The laser pulse was provided by a digitally modulated Omicron PhoxX+638 nm diode laser with a 100 Hz repetition rate. The laser spot size was expanded to cover the active pixel and the continuous wave intensity over the cell pixel area was approximately 550 mWcm$^{-2}$ during the pulse. The preconditioning bias was applied using the data acquisition card and the system was controlled by a custom Labview code.

**Transient ideality factors with $V_{pre}$ = 0 V**

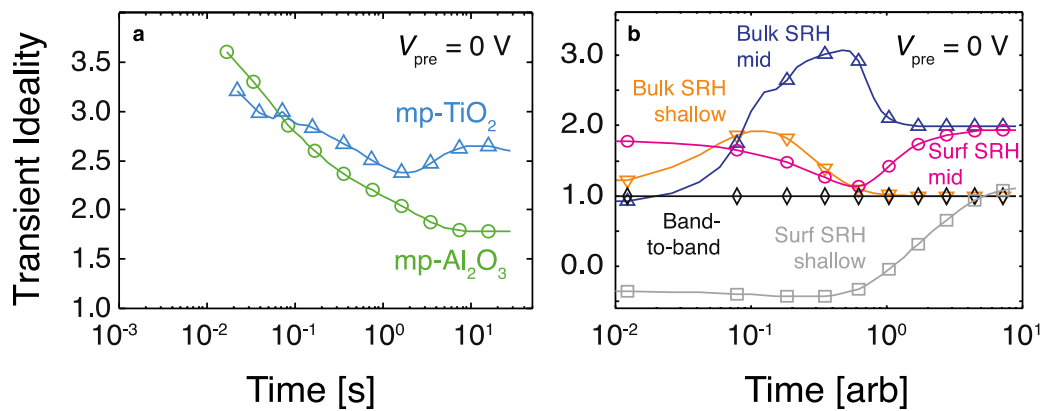

**Figure S9. Measured and simulated evolution of the ideality factor with time following preconditioning at short circuit ($V_{pre}$ = 0 V).** Measured $n_{id}(t)$ for mesoporous TiO$_2$ (blue curve, triangle markers), mesoporous Al$_2$O$_3$ (green curve, circle markers) cells following dark $V_{pre}$ = 0 V for the (**a**) mp- Al$_2$O$_3$ and mp-TiO$_2$ devices). (**b**) The simulated transient ideality factor, $n_{id}(t)$, derived from simulations of $V_{OC}(t)$ with $V_{pre}$ = 0 V, with the following recombination mechanisms: bulk band-to-band recombination (black curve, diamond markers), interfacial SRH recombination *via* shallow traps (grey curve, square markers), interfacial SRH recombination *via* deep traps (pink curve, circle markers), bulk SRH recombination *via* shallow traps (yellow curve, inverted triangle markers), and bulk SRH recombination *via* deep traps (dark blue curve, triangle markers). The distribution of ions immediately following illumination leads to a reverse internal electric field which results in an accumulation of free holes at the n-type interface such that $p > n$ and an accumulation of electrons at the p-type interface, $n > p$ (Figure 5b). Although deeper into the n and p-type contact there is a point where $n = p$ which will also contribute to recombination. This is very different from what would be expected if no mobile ionic charge were present. A consequence is that an instantaneous ideality factor with an intermediate value of $n_{id}(t≈0) ≈ 1.75$ is observed. During the subsequent redistribution of ions the cell tends towards an ideality factor of 2 since $n ≈ p$ right at the interfaces in the final state ($β$ = 2 and $γ$ = 1). However, due to the different rates of ion redistribution which are slower with increasing $φ$ (see $V_{OC}$ transients in both Figure 2b and Figure 2d) the net result is a dip in the transient ideality factor during this process resulting in the observed signature $n_{id}(t)$ seen in (**b**).

**Linearity of simulated transient Suns-$V_{OC}$ curves**

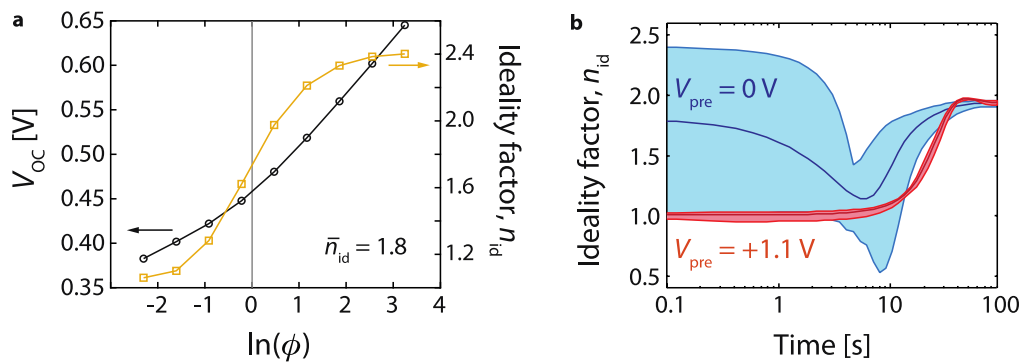

**Figure S10. Example Suns-$V_{OC}$ curve following preconditioning with $V_{pre}$ = 0 V and comparison of the range of idealities following two preconditioning voltages.** (**a**) Example of the non-linearity of the open circuit voltage versus light intensity curve following $V_{pre}$ = 0 V at $t$ = 0.26 s for recombination scheme 3. (**b**) Average Simulated ideality factor $n_{id}$ as a function of time for $V_{pre}$ = 1.1 V (red) and $V_{pre}$ = 0 V (blue) pre-bias conditions. Shaded areas indicate the range of values.

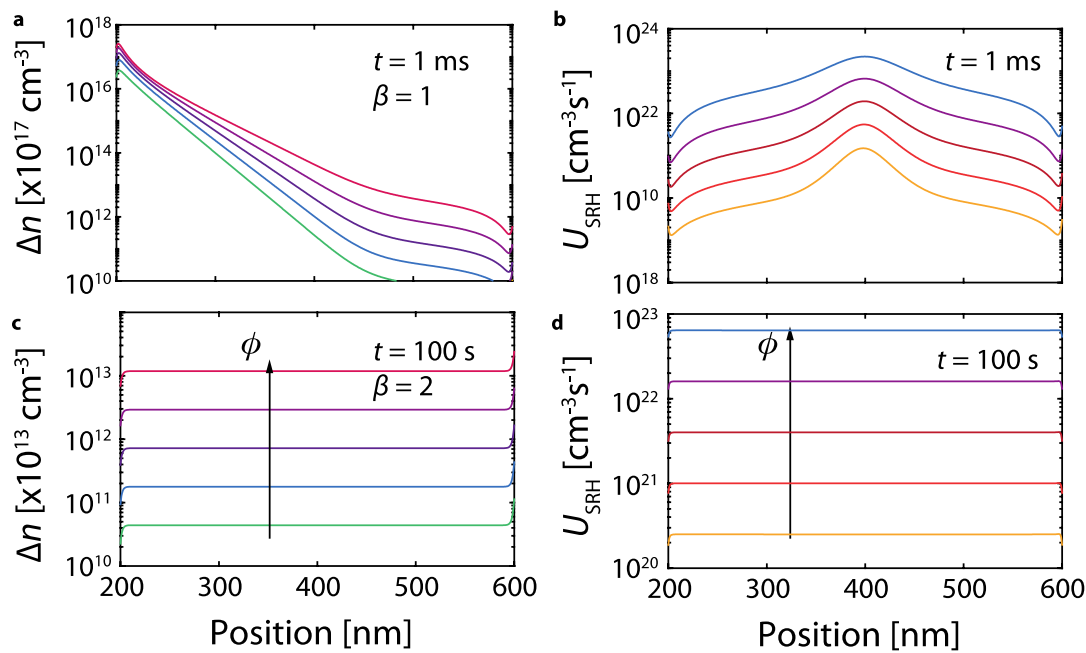

**Figure S11. Change in the excess electron density and recombination profiles following $V_{pre}$ = 0V in a simulated device dominated by bulk SRH recombination.** (**a**) The excess electron density $\Delta n$ and (**b**) Shockley-read-Hall recombination rate $U_{SRH}$ profiles for a device preconditioned with $V_{pre}$ = 0 V and dominated by bulk SRH recombination, directly following illumination ($t$ = 1 ms). The Fermi level splitting at this time is determined by the excess minority accumulation in the space charge regions, leading to $\beta$ = 1. (**c**) $\Delta n$ and (**d**) $U_{SRH}$ profiles following ion migration ($t$ = 100 s). In this steady-state condition, the carrier densities are uniform in the absorber layer, leading to $\beta$ = 2.

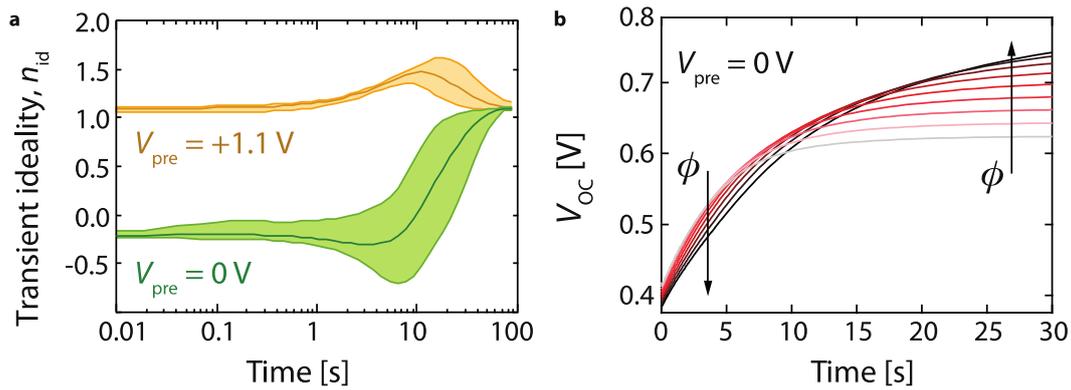

**Figure S12. Simulated transient ideality and $V_{OC}$ transients for a device with high rates of band-to-band recombination in the contact regions.** (**a**) Simulated transient ideality factor for a device dominated by second order band-to-band recombination in the contact regions $V_{pre}$ = +1.3 V (yellow curves) and $V_{pre}$ = 0 V (green curves) pre-bias conditions. The shaded regions indicate the range of values. As the trap energy tends to shallower values, SRH recombination tends to the band-to-band expression, hence these curves are also characteristic of interfacial SRH recombination with shallow traps. (**b**) Transient $V_{OC}$ curves at different light intensities for $V_{pre}$ = 0 V, illustrating the origin of the negative idealities at early times.

**Device Architectures, J-V curves and hysteresis**

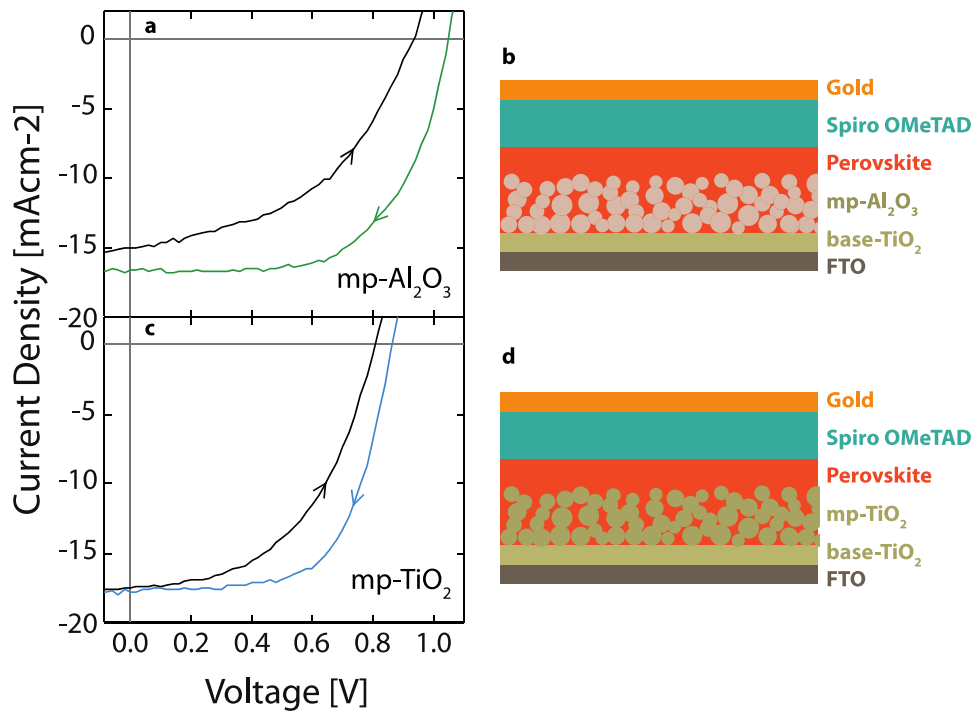

**Figure S13. Influence of scanning directions on *J-V* hysteresis and architectures for the two device types used in the study.** (**a**) The current-voltage (*J-V*) scan for the (**b**) mp-Al$_2$O$_3$ perovskite device used in the study shows a large hysteresis between forward and reverse scans. (**c**) The *J-V* for the (**d**) mp-TiO$_2$ architecture shows a less pronounced hysteresis implying less dependence of the recombination rate on the ionic charge distribution compared to the mp-Al$_2$O$_3$ device. Scan protocol: The scans start from forward bias (FB) to reverse bias (RB) (red lines), or from RB-FB (blue lines) and are measured under AM1.5G simulated solar irradiance of 100 mW cm$^{-2}$ with a scanning rate of 0.15 V s$^{-1}$.

**Optical modelling of devices with transparent conductive adhesive (TCA) electrode**

The normalised optical power density was modelled for a mesoporous $TiO_2$ device stack with a transparent conductive adhesive (TCA) top electrode using a simple Beer-Lambert function such that the light intensity $\varphi$ at position $x$ in each layer falls exponentially in each layer:

$$\varphi(x) = \varphi_0 \exp(-\alpha x)$$

where $\varphi_0$ is the intensity at the start of the layer and $\alpha$ is the absorption coefficient of the material. For the mesoporous $TiO_2$ the optical constants were calculated using the values for pure perovskite and $TiO_2$ phases in proportion 1:1 by volume fraction. The Polyethylene terephthalate (PET) layer of the TCA was assumed to be completely transparent. The optical constants were obtained from references.[33–38]

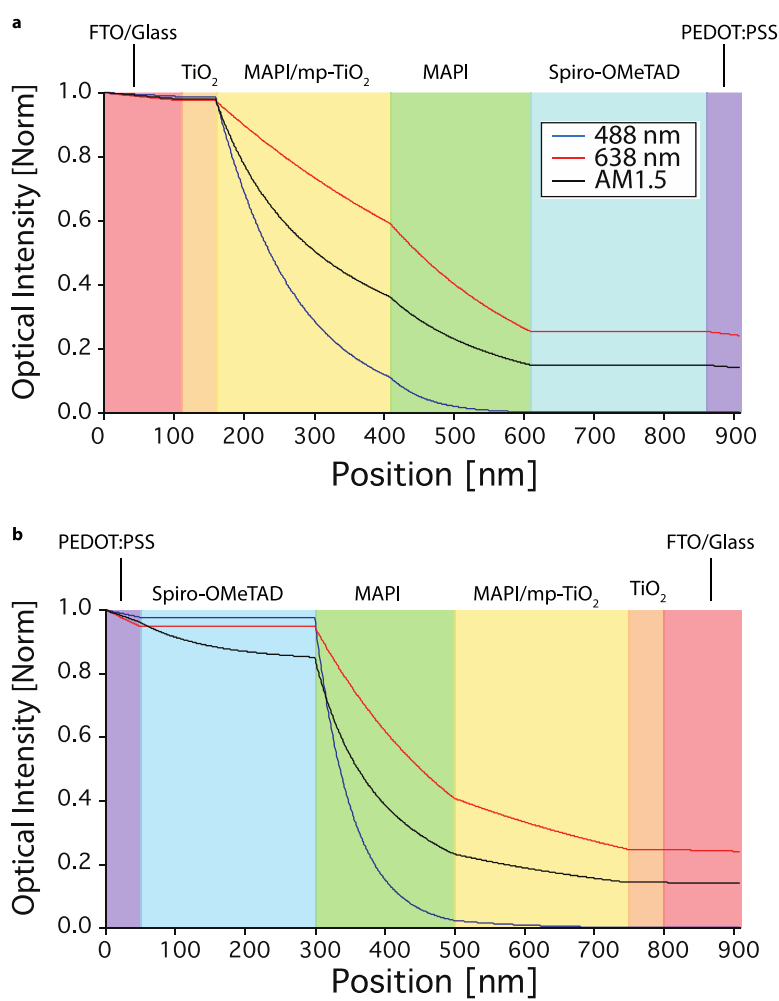

**Figure S14. Optical modelling of the mesoporous $TiO_2$ device with transparent conductive adhesive electrode used in the transient photovoltage measurements.** Optical intensity as a function of position for the mp-$TiO_2$ device stack incorporating the transparent conduction adhesive (TCA) electrode. Illumination from (**a**) the FTO/glass side and (**b**) the TCA side. Blue and red curves indicate 488 nm and 638 nm wavelengths, while the black curve shows the decay of the AM1.5 spectrum integrated from 300 nm to 768 nm.

**Transient photovoltage measurements**

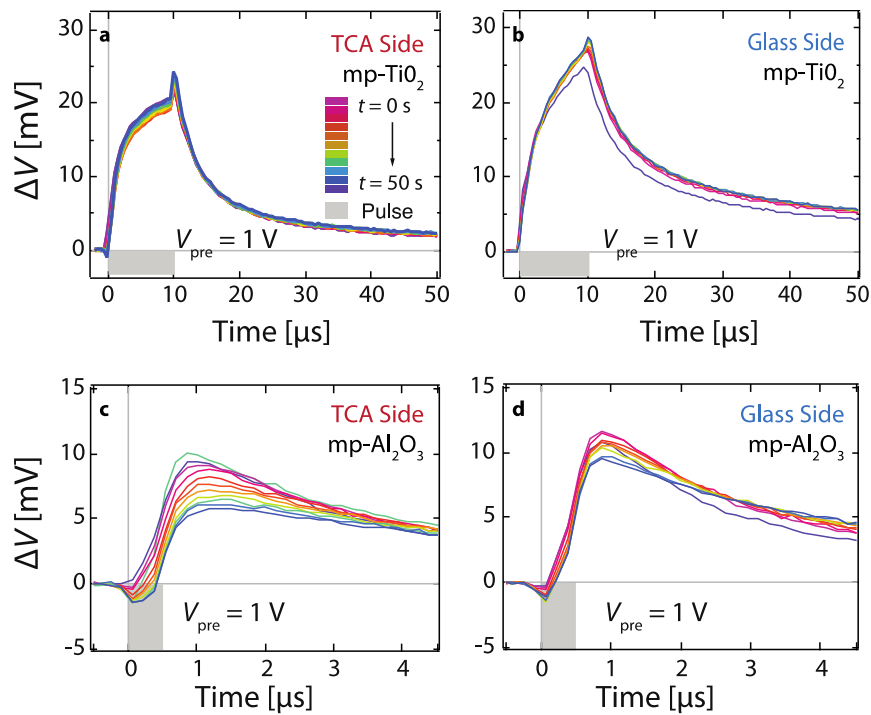

**Figure S15. Transient of the transient photovoltage measurements on bifacial devices following preconditioning at $V_{pre}$ = 1 V.** Transient photovoltage $\Delta V$ following preconditioning at $V_{pre}$ = 1 V in the dark for 60 s, excited by a 10 $\mu$s 488 nm laser pulse (grey shaded region). mp-TiO$_2$ device illuminated through (**a**) TCA and (**b**) glass side show similarly positive signal cf. Figure 6a and Figure 6b. Measurements on the mp-Al$_2$O$_3$ device also show similarly improved magitudes to the signal when illuminated through the (**c**) TCA and (**d**) glass side compared with when the device was preconditioned at $V_{pre}$ = 0 V (Figure 6d and Figure 6e). The reduction in the negative signal indicates that the preconditioning has reduced the strength of the inverted field present at open circuit.

**References**

1. Sah, C. T., Noyce, R. N. & Shockley, W. Carrier Generation and Recombination in P-N Junctions and P-N Junction Characteristics. *Proc. IRE* **45,** 1228–1243 (1957).
2. Koster, L. J. *et al.* Light intensity dependence of open-circuit voltage of polymer: fullerene solar cells. *Appl. Phys. Lett.* **86,** 123509 (2005).
3. Kirchartz, T., Deledalle, F., Tuladhar, P. S., Durrant, J. R. & Nelson, J. On the Differences between Dark and Light Ideality Factor in Polymer:Fullerene Solar Cells. *J. Phys. Chem. Lett.* **4,** 2371–2376 (2013).
4. Calado, P. *et al.* Evidence for ion migration in hybrid perovskite solar cells with minimal hysteresis. *Nat. Commun.* **7,** (2016).
5. Pockett, A. *et al.* Characterization of Planar Lead Halide Perovskite Solar Cells by Impedance Spectroscopy, Open-Circuit Photovoltage Decay, and Intensity-Modulated Photovoltage/Photocurrent Spectroscopy. *J. Phys. Chem. C* **119,** 3456–3465 (2015).
6. Tress, W. *et al.* Interpretation and Evolution of Open-Circuit Voltage, Recombination, Ideality Factor and Subgap Defect States during Reversible Light-Soaking and Irreversible Degradation of Perovskite Solar Cells. *Energy Environ. Sci.* **11,** 151–165 (2017).
7. Sarritzu, V. *et al.* Optical determination of Shockley-Read-Hall and interface recombination currents in hybrid perovskites. *Sci. Rep.* **7,** 44629 (2017).
8. van Reenen, S., Kemerink, M. & Snaith, H. J. Modeling Anomalous Hysteresis in Perovskite Solar Cells. *J. Phys. Chem. Lett.* **6,** 3808–3814 (2015).
9. Pockett, A. & Carnie, M. J. Ionic Influences on Recombination in Perovskite Solar Cells. *ACS Energy Lett.* **2,** 1683–1689 (2017).
10. Belisle, R. A. *et al.* Interpretation of inverted photocurrent transients in organic lead halide perovskite solar cells: proof of the field screening by mobile ions and determination of the space charge layer widths. *Energy Environ. Sci.* (2016). doi:10.1039/C6EE02914K
11. O'Regan, B. C. *et al.* Opto-electronic studies of methylammonium lead iodide perovskite solar cells with mesoporous TiO2; separation of electronic and chemical charge storage, understanding two recombination lifetimes, and the evolution of band offsets during JV hysteresis. *J. Am. Chem. Soc.* **137,** 5087–5099 (2015).
12. Neukom, M. T. *et al.* Why perovskite solar cells with high efficiency show small IV-curve hysteresis. *Sol. energy Mater. Sol. cells* **169,** 159–166 (2017).
13. Calado, P., Barnes, P. R. F., Gelmetti, I., Azzouzi, M. & Hilton, B. Driftfusion. (2017).
14. Shockley, W. & Read, W. T. Statistics of the Recombination of Holes and Electrons. *Phys. Rev.* **87,** 835–842 (1952).
15. Xu, J. *et al.* Perovskite–fullerene hybrid materials suppress hysteresis in planar diodes. *Nat. Commun.* **6,** 7081 (2015).
16. Wojciechowski, K. *et al.* Heterojunction Modification for Highly Efficient Organic-Inorganic Perovskite Solar Cells. *ACS Nano* **8,** 12701–12709 (2014).
17. Wetzelaer, G. A. H., Kuik, M., Lenes, M. & Blom, P. W. M. Origin of the dark-current ideality factor in polymer : fullerene bulk heterojunction solar cells. *Appl. Phys. Lett.* **99,** (2011).
18. Foertig, A., Rauh, J., Dyakonov, V. & Deibel, C. Shockley equation parameters of P3HT:PCBM solar cells determined by transient techniques. *Phys. Rev. B* **115302,** 1–7 (2012).
19. Kirchartz, T., Pieters, B. E., Kirkpatrick, J., Rau, U. & Nelson, J. Recombination via tail states in polythiophene:fullerene solar cells. *Phys. Rev. B* **115209,** 1–13 (2011).
20. Breitenstein, O., Altermatt, P., Ramspeck, K. & Schenk, A. The Origin of Ideality factors n > 2 of Shunts and Surfaces in the Dark I-V Curves of Si Solar Cells. *21st Eur. Photovolt. Sol. Energy Conf.* 625–628 (2006).
21. Kirchartz, T. & Nelson, J. Meaning of reaction orders in polymer:fullerene solar cells. *Phys. Rev. B* **86,** 1–12 (2012).
22. Lee, M. M., Teuscher, J., Miyasaka, T., Murakami, T. N. & Snaith, H. J. Efficient Hybrid Solar Cells Based on Meso-Superstructured Organometal Halide Perovskites. *Science (80-. ).* **338,**

643–647 (2012).
23. Kim, H.-S. *et al.* Lead iodide perovskite sensitized all-solid-state submicron thin film mesoscopic solar cell with efficiency exceeding 9%. *Sci. Rep.* **2,** 591 (2012).
24. Barnes, P. R. F. *et al.* Interpretation of optoelectronic transient and charge extraction measurements in dye-sensitized solar cells. *Adv. Mater.* **25,** 1881–922 (2013).
25. Yao, J. *et al.* Quantifying losses in open-circuit voltage in solution-processable solar cells. *Phys. Rev. Appl.* **4,** 1–10 (2015).
26. Yin, W.-J., Shi, T. & Yan, Y. Unusual defect physics in CH3NH3PbI3 perovskite solar cell absorber. *Appl. Phys. Lett.* **104,** 063903/1-063903/4 (2014).
27. Du, M. Density Functional Calculations of Native Defects in CH3NH3PbI3: Effects of Spin-Orbit Coupling and Self-Interaction Error. *J. Phys. Chem. Lett.* **6,** 1461–1466 (2015).
28. O'Regan, B. C., Durrant, J. R., Sommeling, P. M. & Bakker, N. J. Influence of the TiCl4 Treatment on Nanocrystalline TiO 2 Films in Dye-Sensitized Solar Cells. 2. Charge Density, Band Edge Shifts, and Quantification of Recombination Losses at Short Circuit. *J. Phys. Chem. C* **111,** 14001–14010 (2007).
29. Docampo, P. *et al.* Control of Solid-State Dye-Sensitized Solar Cell Performance by Block-Copolymer-Directed TiO2 Synthesis. *Adv. Funct. Mater.* **20,** 1787–1796 (2010).
30. Correa-Baena, J.-P. *et al.* Identifying and suppressing interfacial recombination to achieve high open-circuit voltage in perovskite solar cells. *Energy Environ. Sci.* **10,** 1207–1212 (2017).
31. Bryant, D. *et al.* A Transparent Conductive Adhesive Laminate Electrode for High Efficiency Organic-Inorganic Lead Halide Perovskite Solar Cells. *Adv. Mater.* **26,** 7499–7504 (2014).
32. Domanski, K. *et al.* Migration of cations induces reversible performance losses over day/night cycling in perovskite solar cells. *Energy Environ. Sci.* **10,** 604–613 (2017).
33. Leguy, A. M. A. *et al.* Experimental and theoretical optical properties of methylammonium lead halide perovskites. *Nanoscale* **8,** 6317–6327 (2016).
34. Schulz, P. *et al.* Electronic Level Alignment in Inverted Organometal Perovskite Solar Cells. *Adv. Mater. Interfaces* **2,** 1–5 (2015).
35. Onoda-Yamamuro, N., Matsuo, T. & Suga, H. Dielectric study of CH3NH3PbX3 (X = Cl, Br, I). *J. Phys. Chem. Solids* **53,** 935–939 (1992).
36. Eames, C. *et al.* Ionic transport in hybrid lead iodide perovskite solar cells. *Nat. Commun.* **6,** 7497 (2015).
37. Leijtens, T. *et al.* Electronic Properties of Meso-Superstructured and Planar Organometal Halide Perovskite Films: Charge Trapping, Photodoping, and Carrier Mobility. *ACS Nano* **8,** 7147–7155 (2014).
38. Burkhard, G. F., Hoke, E. T. & McGehee, M. D. Transfer Matrix Optical Modeling. (2011).